\documentclass[trackchanges,twocolumn]{aastex701}

\begin{document}

\title{Searching for Type Ia Supernovae in the Dark Energy Spectroscopic Instrument}

\correspondingauthor{Xiaoyu Zhuang, Xu Kong}

\author[0009-0002-3759-9275,sname=Zhuang,gname=Xiaoyu]{Xiaoyu Zhuang}
\affiliation{Department of Astronomy, University of Science and Technology of China, Hefei 230026, China}
\affiliation{School of Astronomy and Space Science, University of Science and Technology of China, Hefei 230026, China}
\email[show]{zxy0543@mail.ustc.edu.cn}

\author[0009-0000-2386-5881,gname=Song, sname=Ran]{Song Ran} 
\affiliation{Department of Astronomy, University of Science and Technology of China, Hefei 230026, China}
\affiliation{School of Astronomy and Space Science, University of Science and Technology of China, Hefei 230026, China}
\email{599631458@qq.com}

\author[gname=Dezheng,sname=Meng]{Dezheng Meng}
\affiliation{Department of Astronomy, University of Science and Technology of China, Hefei 230026, China}
\affiliation{School of Astronomy and Space Science, University of Science and Technology of China, Hefei 230026, China}
\email{dezhengmeng@mail.ustc.edu.cn}

\author[0000-0003-0230-4596,gname=Weiyu,sname=Ding]{Weiyu Ding}
\affiliation{Department of Astronomy, University of Science and Technology of China, Hefei 230026, China}
\affiliation{School of Astronomy and Space Science, University of Science and Technology of China, Hefei 230026, China}
\email{dingwy@mail.ustc.edu.cn}

\author[0009-0006-4622-1417,gname=Ranfang,sname=Zheng]{Ranfang Zheng}
\affiliation{Department of Astronomy, University of Science and Technology of China, Hefei 230026, China}
\affiliation{School of Astronomy and Space Science, University of Science and Technology of China, Hefei 230026, China}
\email{rfzheng@mail.ustc.edu.cn}

\author[0000-0002-6873-8779,gname=Yao,sname=Yao]{Yao Yao}
\affiliation{Department of Astronomy, University of Science and Technology of China, Hefei 230026, China}
\affiliation{School of Astronomy and Space Science, University of Science and Technology of China, Hefei 230026, China}
\email{yaoyao97@mail.ustc.edu.cn}

\author[0000-0003-4959-1625,gname=Zheyu,sname=Lin]{Zheyu Lin}
\affiliation{Department of Astronomy, University of Science and Technology of China, Hefei 230026, China}
\affiliation{School of Astronomy and Space Science, University of Science and Technology of China, Hefei 230026, China}
\email{linzheyu@mail.ustc.edu.cn}

\author[0000-0003-4122-6949,gname=Bingxue,sname=Fu]{Bingxue Fu}
\affiliation{Department of Astronomy, University of Science and Technology of China, Hefei 230026, China}
\affiliation{School of Astronomy and Space Science, University of Science and Technology of China, Hefei 230026, China}
\email{fbx001128@mail.ustc.edu.cn}

\author[0000-0001-9472-2052,gname=Fujia,sname=Li]{Fujia Li}
\affiliation{Department of Astronomy, University of Science and Technology of China, Hefei 230026, China}
\affiliation{School of Astronomy and Space Science, University of Science and Technology of China, Hefei 230026, China}
\email{lifujia@mail.ustc.edu.cn}

\author[gname=Zelin,sname=Xu]{Zelin Xu}
\affiliation{Department of Astronomy, University of Science and Technology of China, Hefei 230026, China}
\affiliation{School of Astronomy and Space Science, University of Science and Technology of China, Hefei 230026, China}
\email{sa21022025@mail.ustc.edu.cn}

\author[0000-0002-0846-7591,gname=Jie,sname=Song]{Jie Song}
\affiliation{Department of Astronomy, University of Science and Technology of China, Hefei 230026, China}
\affiliation{School of Astronomy and Space Science, University of Science and Technology of China, Hefei 230026, China}
\email{jiesong@mail.ustc.edu.cn}

\author[0000-0002-7660-2273,gname=Xu,sname=Kong]{Xu Kong}
\affiliation{Department of Astronomy, University of Science and Technology of China, Hefei 230026, China}
\affiliation{School of Astronomy and Space Science, University of Science and Technology of China, Hefei 230026, China}
\affiliation{Institute of Deep Space Sciences, Deep Space Exploration Laboratory, Hefei 230026, China}
\email[show]{xkong@ustc.edu.cn}

\begin{abstract}

With the development of large‑scale photometric surveys, an increasing number of supernova candidates are being discovered, leading to a rapidly growing demand for supernova spectra. In addition to equipping photometric surveys with follow-up spectroscopic facilities, archival spectra from large multi-object spectroscopic surveys can be mined to provide spectroscopic classifications for candidates and to find supernovae missed by previous surveys. In this work, we combine Principal Component Analysis (PCA),  the Local Outlier Factor (LOF) algorithm, and the supernova classification tool \texttt{SNID} to search for Type Ia supernovae among 1,757,303 galaxy spectra from the Dark Energy Spectroscopic Instrument (DESI) Data Release 1 (DR1). We finally obtain 247 Type Ia supernovae and 17 supernovae of other types. Among these, 202 supernovae lack classification records in the Transient Name Server (TNS) and represent newly identified SNe. These results demonstrate the potential of multi-object spectroscopic surveys to supplement supernova samples, particularly for transients missed by traditional photometric surveys. 

\end{abstract}



\section{Introduction} 
A supernova (SN) is a violent explosion that occurs at the end of a massive star's life or results from a runaway thermonuclear fusion in a white dwarf. The former is referred to as a core‑collapse supernova, while the latter is known as a thermonuclear supernova (also known as Type Ia supernova; SN Ia). Supernovae are not only crucial stages in stellar evolution but also pivotal events in the dynamical evolution of the cosmos, the chemical evolution of galaxies, and the formation of neutron stars and black holes. Furthermore, they serve as ``standard candles" for astronomical distance measurements \citep{Filippenko_1997}, holding significant research value in astrophysics.

The most commonly used method for SN searches is based on image subtraction \citep[e.g.][]{Perlmutter_1995,Bailey_2007,Hu_2022}. This technique involves taking images of the same sky region at different epochs, comparing them with template images to identify objects with brightness variations, and subsequently measuring their light curves to classify the candidates. In earlier periods (around 1990–2010), dedicated telescopes were employed to monitor known galaxies or specific sky regions, such as the Lick Observatory Supernova Search \citep[LOSS,][]{Li_2000}, the Palomar Transient Factory \citep[PTF,][]{Law_2009}, and the Beijing Astronomical Observatory Supernova Survey \citep[BAOSS,][]{Qiu_Hu_Li_2001}. With the development of large-aperture telescopes and expansive sky survey projects, an increasing number of SNe are being discovered. Large sky survey projects with multiple scientific goals, such as the Panoramic Survey Telescope and Rapid Response System \citep[Pan‑STARRS,][]{Chambers_2016}, the Zwicky Transient Facility \citep[ZTF,][]{Bellm_2019}, and the Wide Field Survey Telescope \citep[WFST,][]{WFST_2023}, exemplify this trend.

Although the image‑subtraction method is efficient for identifying SNe, significant gaps remain in the spectroscopic follow-up of these transients. To date, approximately 180,000 transients have been reported to TNS \citep{Gal2021TNS} since 2016 January 1, but only about 12\% of these transients have spectroscopic classifications. Despite many advancements in the photometric classification of SNe in recent years \citep[e.g.][]{Lochner_2016,Charnock_2017,Garg_2025}, the precise classification still relies on characteristic absorption features in their spectra. Furthermore, spectra are necessary for detailed physical studies regarding the evolution of SNe and their interactions with host galaxies. For example, early spectra of Type IIP SN 2024jlf showed weak flash ionization features that disappeared after less than two days, indicating that the progenitor exhibited enhanced mass-loss before explosion \citep{Rehemtulla_2025}. Spectral analysis of Type Ia SNe and their host galaxies reveals that higher-velocity SNe Ia tend to explode in more massive galaxies and in the inner regions of their hosts \citep{Pan_2015}. Those studies would be impossible without sufficient spectra. 

Unfortunately, many photometric surveys that dominate SN discovery lack dedicated spectroscopic follow‑up capabilities. Consequently, obtaining spectra for newly discovered SNe often requires applying for targeted time on other spectroscopic instruments. This leads to a time delay between the discovery of a candidate and the acquisition of its follow‑up spectrum, potentially missing the optimal period for spectroscopic observation and the short-lived features present in the early stages. Even if the follow-up spectroscopic instruments are available, the observation of a target's spectrum requires more integration time than imaging, leaving a large portion of transients without spectroscopic confirmation. 

In addition to photometric surveys, numerous spectroscopic surveys operate worldwide, such as the Sloan Digital Sky Survey \citep[SDSS,][]{SDSS_2000}, the Dark Energy Spectroscopic Instrument \citep[DESI,][]{DESI_2022}, and the Large Sky Area Multi‑Object Fiber Spectroscopic Telescope \citep[LAMOST,][]{LAMOST_2012}. Although their efficiency in discovering transients is lower than that of photometric surveys, their wide coverage area and vast spectroscopic data make it highly probable that they have occasionally captured spectra of SNe exploding within targeted galaxies. These ``accidentally" captured SN spectra can not only provide information at different epochs for currently known SNe, but also serve as supplement for candidates missing timely follow-up observations. Furthermore, those spectroscopic surveys have the potential to discover SNe missed by photometric surveys. 

Numerous attempts have already been made to search for SNe within spectroscopic surveys. \cite{Madgwick_2003} first proposed a PCA-based method to eliminate the contamination of host galaxies and used the residual spectra to seek SNe in SDSS-DR1. \cite{Tu_2010} improved upon this method by introducing SN eigenspectra and implemented it on SDSS-DR7. \cite{Krughoff_2011} further refined candidate selection by employing a non-parametric Bayesian classifier based on decomposed spectra, applying their method to SDSS-DR5. \cite{Graur_2013} enhanced the reliability of candidates by incorporating $\chi^2$ fitting criteria and spectral classification criteria into the decomposition approach, validating their method on SDSS-DR7. Subsequently, \cite{Graur_2015} applied the same method to SDSS-DR9. With the advancement of artificial intelligence, recent years have seen attempts to use machine learning methods for SN searches, such as the convolutional neural network DASH developed by \cite{Muthukrishna_2019}, and the Recurrent Neural Network SNIascore developed by \cite{SNIascore}.

DESI is a robotically-actuated, fiber-fed spectrograph installed at prime focus on the 4 m Mayall telescope in Kitt Peak, Arizona \citep{DESI_2016a,DESI_2016b}. It commenced regular observations in 2021 and has recently released Data Release 1 \citep[DR1,][]{DESIDR1}. DESI can obtain 5,000 spectra in the wavelength range of 3600 \AA\ to 9800 \AA\ in a 3$^\circ$ diameter field of view in a single exposure, resulting in an average output of over 2 million spectra per month. Over its five-year survey, DESI will observe 40 million galaxies and quasars as well as over 8 million stars in a sky area of approximately 14,000 $deg^2$. Given its extensive coverage and massive data volume, DESI has a high probability of capturing spectra of SNe. Furthermore, the Mayall telescope enables the detection of galaxies at higher redshifts or with lower luminosities, potentially revealing faint SNe that were undetectable by previous surveys.

As DESI data are relatively new, few studies have utilized them for SN searches. Existing work primarily involves cross-matching transients discovered by other surveys with DESI data to provide spectra for the transients and their host galaxies \citep{Soumagnac_2024}. Such works are limited by the detection capability of other photometric surveys, and thus cannot fully utilize the deeper detection limit of DESI. Moreover, the large data volume of DESI makes manual SN searches prohibitively time-consuming. Therefore, building upon the methodologies established by previous researchers, we conducted an almost automated SN search in DESI DR1 with no transient catalogs incorporated in the data reduction.

The organization of this paper is as follows: Section \ref{sec:data} introduces the DESI DR1 data and the preprocessing methods. Section \ref{sec:method} details the search method and the vital parameters used. Section \ref{sec:result} presents the SNe discovered and discusses the cross-match result between the SNe in TNS and our sample. Finally, Section \ref{sec:sum} summarizes the findings of this work and provides an outlook for future research.

\section{Data Reduction and Preprocessing} \label{sec:data}
\subsection{DESI DR1 sample selection}
The DESI DR1 spectra are observed through fibers installed on 10 identical spectrographs. Each fiber has a diameter of 107 $\mu$m, corresponding to 1.4 arcsecs on the sky \citep{Poppett_2020}. The spectra cover three wavelength bands: B (3600–5800 \AA), R (5760–7620 \AA), and Z (7520–9824 \AA), with a wavelength interval of 0.8 \AA. Each band has a different resolution, with a minimal resolution of $R\approx2000$. 

DESI DR1 comprises two parts: the spectra from the Main Survey taken between May 2021 and June 2022, and the spectra taken as part of DESI Survey Validation from December 2020 to June 2021. The latter data were originally released as part of the Early Data Release (EDR) but were reprocessed using the same pipeline as the Main Survey, correcting many errors in redshift measurement or spectrum type determination. 

DESI's primary targets include the Bright Galaxy Survey (BGS) galaxies, Emission Line Galaxies (ELGs), Luminous Red Galaxies (LRGs), Quasars (QSOs), and Milky Way Survey (MWS) and backup program stars \citep{Myers_2023}. Galaxy spectra in DESI mainly belong to the first three categories. Galaxies belonging to BGS usually have $0<z<0.6$ \citep{Hahn_2023}, and are observed during `bright' time (when the moon is up). Galaxies belonging to ELG usually have $0.6<z<1.6$ \citep{Raichoor_2023}, showing distinct emission lines from star formation activities and hot young stars. Galaxies belonging to LRG usually have $0.4<z<1.1$ \citep{Zhou_2023}, with prominent 4000 \AA\ break and older stellar population. Both ELGs and LRGs are observed during `dark' time (when the moon is down), and some of them can appear in lower redshift. Target classifications (e.g., BGS, ELG) are initially selected based on DESI Legacy Surveys imaging photometry; however, the ultimate classification depends on the resulting spectrum.

 DESI spectra are classified into three primary types by the \texttt{redrock}\footnote{\url{https://github.com/desihub/redrock/}} fit: GALAXY, QSO, and STAR. Since SNe generally occur within galaxies, we utilized spectra classified as ``GALAXY''. In DR1, there are 22,720,067 galaxy spectra. To ensure that SN features fall within the selected wavelength range, we adopted the redshift range of 0 to 0.25. To maintain data quality, we required that the median signal-to-noise ratio in the R band (MEDIAN\_COADD\_SNR\_R) to be greater than 5, and that valid data exist across all three bands. After applying these criteria, the number of the remaining galaxy spectra reduced to 1,757,303. Considering computational memory constraints and the search efficiency, we divided these spectra into 176 groups, each containing 10,000 spectra. 
 
\subsection{Preprocessing}
Before searching, we performed the following preprocessing steps on the spectra. First, the spectrum is masked according to the mask array. Then, the spectrum is corrected for Galactic dust extinction according to \cite{Cardelli_1989} and the E(B-V) value provided by DESI. Next, the spectrum is smoothed via Gaussian convolution and shifted to the rest frame. A linear interpolation is then performed to resample the spectrum onto a wavelength range of [3600 \AA, 7200 \AA] with a 2.4 \AA\ interval. This interpolation retains primary spectral features while reducing data dimensionality. Finally, the spectrum is normalized by its median flux to mitigate the influence of strong emission lines on the prominence of SN feature. 

\section{Method} \label{sec:method}
Our searching method primarily follows the approach employed by \cite{Tu_2010}, which can be summarized into the following steps:
\begin{enumerate}
    \item Use PCA to extract the eigenspectra of galaxy and supernova templates
    \item Decompose the input spectra into galaxy and supernova components to construct the SuperNova Statistical Characterization Vector (SNSCV).
    \item Perform outlier detection (specifically LOF) based on the SNSCV to screen for candidate spectra.
    \item Classify the candidate spectra using supernova spectrum classification software.
    \item Conduct visual inspection.
\end{enumerate}
This section details several crucial modifications we made to the searching method. 

\subsection{Candidate selection}
Potential candidates containing SNe are selected after the first three steps. To construct the SNSCV, PCA is employed to extract the eigenspectra from  standard galaxy and supernova spectral templates. In classic PCA, finding the principal components (PCs) is equivalent to maximizing the variance of the data in a specific axis. That is to maximize
\begin{equation}
    Var(x)=\frac1mw^T(\sum_{i=1}^mx_ix_i^T)w
    \label{Varx}
\end{equation}
under the constraint condition $w^Tw=1$, where $w$ is the direction vector of an axis and $x_i$ is the data vector after subtracting the mean and normalizing. This can be solved using the method of Lagrange multipliers. 

$Var(x)$ reaches its maximum value when $w$ is the eigenvector of the covariance matrix $C=\sum_{i=1}^mx_ix_i^T$. Eigenvector of $C$ can be obtained via Singular Value Decomposition (SVD). The larger the eigenvalue, the greater the contribution of the variance along the corresponding eigenvector to the total variance, and thus the more important that PC is.

If the input data are not mean-centered, Equation~(\ref{Varx}) no longer represents variance but instead includes information about the overall offset of the data relative to the origin. The SVD can still be performed, but the eigenvalues in this case indicate the importance of the eigenvectors in describing the absolute data distribution rather than variance. Since galaxy spectra and supernova spectra occupy markedly different regions in absolute feature space, retaining this offset information is advantageous for distinguishing between the two types of spectra. As the templates do not provide noise information, we employ classic PCA without assigning weights to the template data. 

After getting $m$ galaxy eigenspectra \{$g_i$\} and $n$ supernova eigenspectra \{$s_i$\}, the origin spectrum $D$ can theoretically be expressed as
\begin{equation}
    D=G+S\approx \sum_{i=1}^m a_{i}g_{i}+\sum_{i=1}^n a_{m+i}s_{i}
    \label{component}
\end{equation}
where $G$ is the galaxy component and $S$ is the supernova component. The coefficients preceding \{$s_i$\}, namely
\[(a_{m+1},a_{m+2},...,a_{m+n})\]
are defined as the SNSCV, which characterizes the supernova component of the spectrum.  

Although SNe happen in galaxies all the time, many are too faint to be detected or explode when their host galaxies are not actively being observed by DESI. Thus, compared with galaxy spectra, accidentally captured SN spectra are rare, allowing them to be treated as outliers within the bulk sample. 

The outlier detection method we employ is the Local Outlier Factor (LOF), which is based on local density \citep{LOF}. An object's $LOF$ is the ratio of the local densities of its neighbors and itself: 
\begin{equation}
LOF_k(P)=(\frac{\sum_{O\in N_k(P)}lrd_k(O)}{|N_k(P)|})/lrd_k(P)
\end{equation}
where $lrd(P)$ indicates the local density of the object P, $N_k(P)$ is P's neighbors, $|N_k(P)|$ is the number of objects in $N_k(P)$, and parameter $k$ suggests that the neighbors of the object extend as far as its k-th nearest neighbor(s), which influences the distance calculation of two objects and thus the local density calculation. A $LOF$ close to 1 indicates that the object belongs to the same cluster as its neighborhood. In contrast, a $LOF$ significantly greater than 1 suggests that the local density of the object is much smaller than that of its neighbors, marking it as an outlier. 

Given that LOF requires calculating pairwise distances between all points, applying LOF to large datasets  composed of high-dimensional vectors can be extremely time-consuming. Subsequently, numerous studies have proposed optimization methods to accelerate the LOF computation \citep[e.g.][]{Abhaya_2022}. We utilize the LOF function\footnote{\url{https://scikit-learn.org/stable/modules/generated/sklearn.neighbors.LocalOutlierFactor.html}} from the Python module \texttt{scikit-learn} \citep{scikit-learn}, which incorporates optimized methods for LOF, significantly improving computational speed.

After calculating the $LOF$ based on their SNSCV, the spectra are sorted in descending order based on their $LOF$. Those top-ranked spectra are considered as potential candidates containing supernovae.

\subsection{Specific parameter selection}
The quality of spectrum decomposition and the 
efficiency of outlier detection both depend on templates. 

The development of galaxy spectral templates has a relatively long history, and many previous studies have utilized PCA to extract galaxy eigenspectra and studied common characteristics of galaxies, such as \cite{Yip_2004}, who performed PCA on about 170,000 galaxy spectra in SDSS. DESI research group also employs PCA to construct eigenspectra for redshift estimation \citep{Bailey_2023}. The galaxy templates they used are generated from stellar population synthesis and emissionline modeling of galaxies at $0<z<1.5$, consisting of 10,000 ELGs, 5,000 LRGs and 5,000 BGS galaxies. They also provide the source code for generating these templates and eigenspectra on desihub\footnote{\url{https://github.com/desihub/redrock/blob/main/bin/rrgaltemplate}}. 

As our galaxy sample falls within $0<z<0.25$, the number of BGS galaxies is far greater than that of ELGs and LRGs. Therefore, we adjusted the proportion of ELGs, LRGs, and BGS galaxies to 1:1:2. We also modified the wavelength range and interval of the synthetic galaxy spectra to match our preprocessing interpolation, namely $\lambda=3600-7200$\AA, $\Delta\lambda=2.4$\AA. The final number of eigenspectra used is 10, with the sum of their corresponding eigenvalues accounting for over 0.99 of the total eigenvalue sum.

Although SN spectral templates also have a development history spanning over two decades, limited by the quantity of high-quality spectra and their diversity, only Type Ia SN spectral templates are relatively comprehensive and mature to date. Early SN Ia templates were constructed mostly by calculating the weighted mean spectra (or weighted mean features) of a carefully processed spectral library and interpolating it to a uniformly spaced time series, such as \cite{Nugent_2002} and \cite{Hsiao_2007}. Later templates were mostly given by an empirical model trained with real observations, incorporating methods like Expectation Maximization Factor Analysis (EMFA) and Gaussian process, e.g., SALT2/3 \citep{Betoule_2014, Taylor_2021, Taylor_2023, Kenworthy_2021}, SNEMO \citep{Saunders_2018} and SUGAR \citep{L_get_2020}. Apart from those `synthetic' templates, there also exists high-quality SN spectral banks that can directly be used for PCA despite the lack of certain epoch, such as the Superfit bank\footnote{\url{https://www.wiserep.org/sites/default/files/supyfit_bank.zip}}\citep{NGSF}. 

To determine the optimal templates, we constructed a test set comprising 9900 synthetic galaxy spectra (generated using \texttt{desisim}\footnote{\url{https://github.com/desihub/desisim}}) and 100 Type Ia SN spectra from the Open Supernova Catalog \citep[OSC,][]{Guillochon_2017} to simulate the search process with fixed galaxy eigenspectra. Most templates were accessed via \texttt{SNCosmo}\footnote{\url{https://sncosmo.readthedocs.io/en/stable/source-list.html}}\citep{barbary_2025_15019859}. As for Superfit, we did PCA on its Ia-norm library, which contains 89 spectra of 13 Ia-norm SNe. We define the ``accuracy'' of a template as the proportion of SN spectra found in the top 1\% spectra after ranked by $LOF$ in descending order. We measure the ``completeness'' of a template by how many spectra it must go through to find all SNe, presented as percentage ranking of the SN ranked last. 

\begin{figure}[htb!]
    \centering
    \gridline{
        \fig{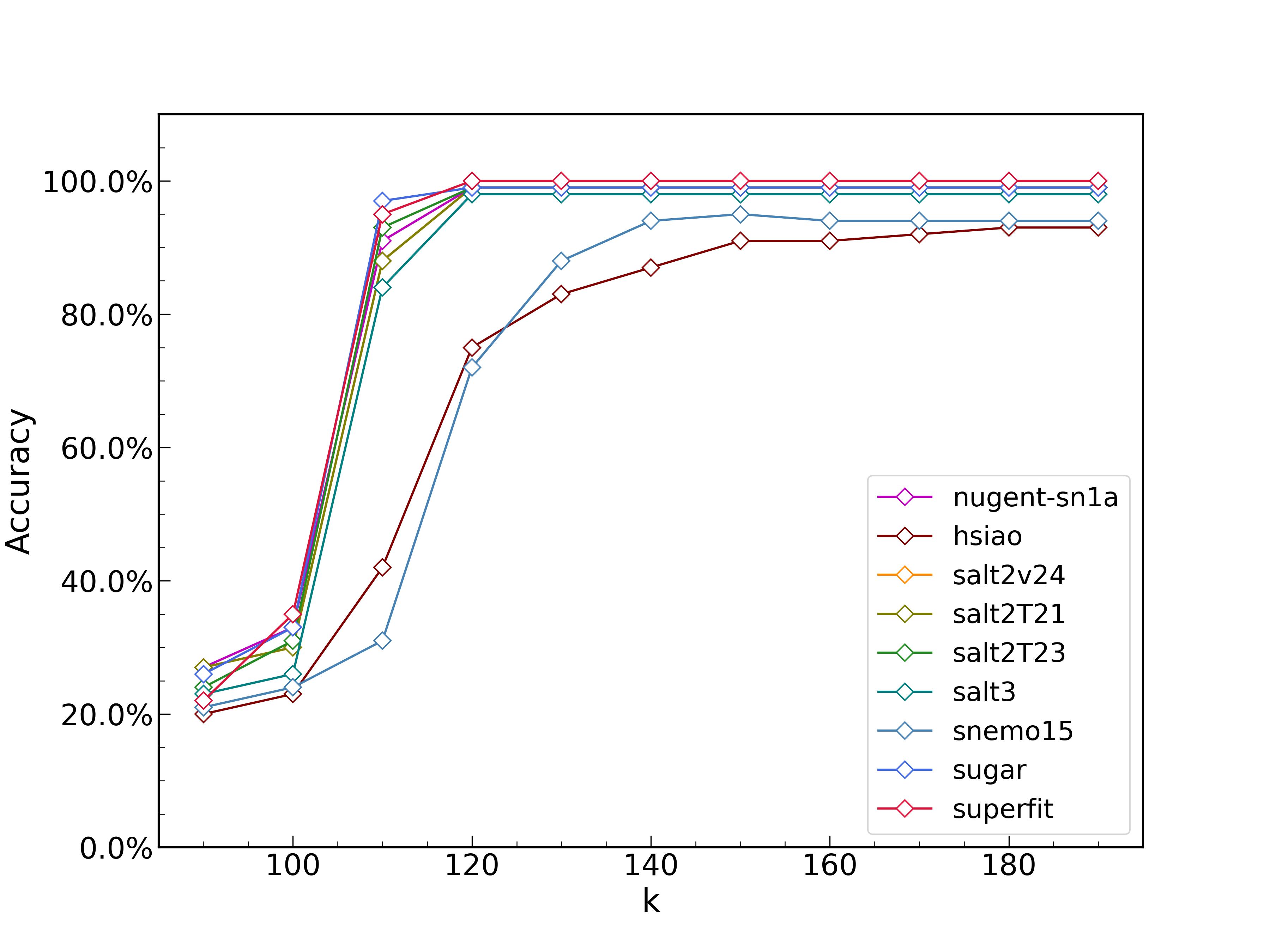}{0.46\textwidth}{(a) The ``accuracy'' of each template on the test set}
    }
    \gridline{
        \fig{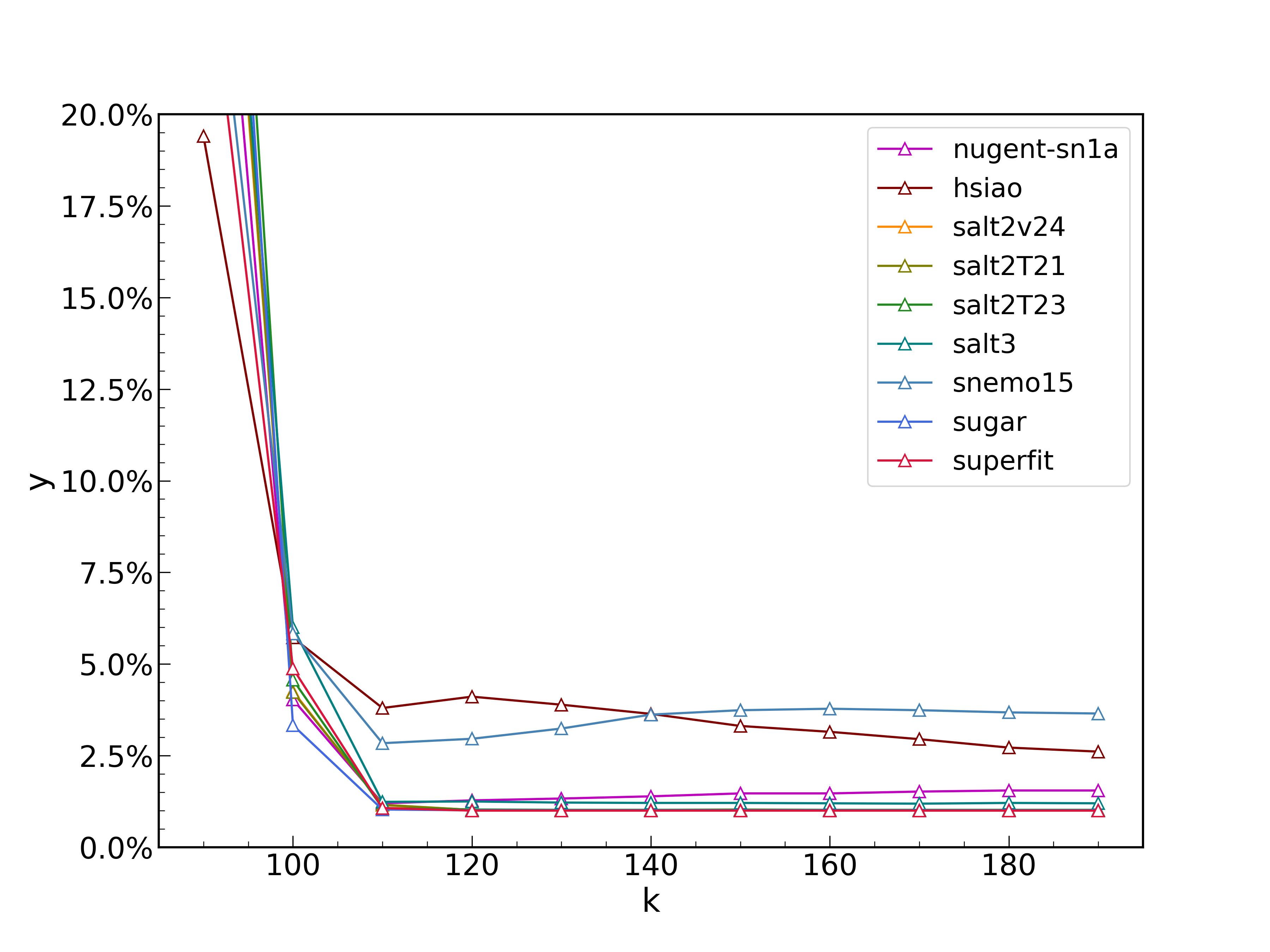}{0.46\textwidth}{(b) The ``completeness'' of each template on the test set}
    }
    \caption{Different SN templates' performance on varying k with fixed galaxy templates, where k specifies the number of neighbors involved in $LOF$ calculation. The test set is composed of 9900 simulated DESI galaxy spectra and 100 Type Ia SN spectra from OSC. ``Accuracy'' refers to the proportion of SN spectra found in the top 1\% spectra after ranked by LOF in descending order. ``y'' refers to the percentage ranking of the last-ranked SN. It can be seen that the performance of templates improves significantly when k is greater than the number of SNe and tends to stabilize when k continues to increase.}
    \label{fig:temptest}
\end{figure}

The test result can be seen in Figure~\ref{fig:temptest}. Ultimately the SUGAR templates demonstrated the best performance, with higher accuracy and less spectra to find all SNe. We extracted eigenspectra from 40 SUGAR template spectra within the age range of $-9$ to $30$ days. The final number of eigenspectra used is 7, with the sum of their corresponding eigenvalues accounting for over 0.99 of the total eigenvalue sum. Other templates were only used for testing purposes and only the SUGAR templates were used in the final search. 

The parameter $k$ matters a lot in LOF calculation. A too small $k$ can't distinguish outliers from other cluster objects, while a too large $k$ will increase the time consumption and storage burden. Therefore, we used the same test set as the templates selection to find a suitable $k$. 

According to Figure~\ref{fig:temptest}, it is necessary to set $k>100$ when a group of 10,000 spectra contains 100 supernovae. When $k > 100$, the template's performance stabilizes and no longer shows significant variation with further increases in $k$. As the actual number of SNe in a real group is even lower, setting $k = 100$ is a reasonable choice. To ensure the completeness, the top 1.5\% of the sorted spectra are selected as candidates. 

\subsection{Classification}\label{subsec:class}
Currently the most widely used supernova spectrum classification software is SuperNova IDentification Code \citep[\texttt{SNID},][]{Blondin_2007}, which classifies spectrum based on template correlation techniques. 

\texttt{SNID} has a built-in template bank containing processed spectra of all types of supernovae and other celestial bodies with brightness variations (like AGN, LBV). The bank is now updated to ``Super-SNID” \citep{Magill_2025}, containing 5760 spectra of 768 celestial bodies. 

After a series of preprocessing steps that reduces the influence of intrinsic color and noises, \texttt{SNID} runs a first set of correlations to estimate an initial redshift guess, then runs a second set of correlations with the input and template spectra trimmed to match at this redshift. If the redshift information is provided, the first set of correlations is skipped.  

\texttt{SNID} uses $rlap=r\times lap$ to quantify the correlation, where $r$ quantifies the significance of the correlation peak in the convolution of the input spectrum with the shifted template \citep{Tonry_1979} and $lap$ is the wavelength overlap range between them \citep{Blondin_2007}. The higher the rlap, the better the correlation. 

Because the template spectra in \texttt{SNID} exhibit obvious SN features, the contamination of the host galaxy can be ignored. However, in real datasets, SNe are not always brighter than their hosts. Many DESI fibers are centered on the nuclear regions of galaxies, resulting in substantial host contamination. Narrow galactic emission lines significantly influences classification accuracy, affecting both redshift estimation and $rlap$ calculation. Therefore, we used the residual spectrum
\begin{equation}
    Res = D-\sum_ia_ig_i
\end{equation}
as the input spectrum of \texttt{SNID}, which can reduce the influence of narrow emission lines to some extent. The definition of the letters is the same as Equation \ref{component}. The SN component $\sum_ia_{m+i}s_i$ was not used as the input spectrum because it tends to be biased toward features of Type Ia SNe, which may lead to a higher rate of misclassification.

DESI's redshift estimation pipeline may fit a wrong redshift to a spectrum containing SN features, as the broad lines produced by SN are not common for galaxy templates \citep[see the example of SN 2021zfs in][]{Davison_2025}. Considering this, we did not force the initial redshift guess of \texttt{SNID}.  

After \texttt{SNID} finished the classification of all candidates, we first ran an automatic selection based on the following criteria: at least one of the top 10 matched templates is flagged with ``good", and the spectral type of the best-matching template belongs to SN; the difference between the redshift estimated by \texttt{SNID} and the redshift provided by DESI is within 0.05, which for the residual spectrum requires $|z_{SNID}| < 0.05$; but if the best-matching template has $rlap>10$, the redshift limitation is waived.

The remaining candidates are then proceed to further visual inspection. 

\section{Result} \label{sec:result}

The method described in Section \ref{sec:method} was applied to the 1,757,303 galaxy spectra from DESI DR1 described in Section \ref{sec:data}. 

Simultaneously, a cross-match between this dataset and SNe recorded in TNS between 2020 November 30 and 2022 July 1 were performed in order to identify known SNe and evaluate the completeness of our method. We utilized a matching radius of 5 arcsecs and constrained the difference between the DESI redshift and the TNS redshift to $\Delta z<0.01$. Furthermore, we required that the difference between the DESI observation date and the SN discovery date in TNS fall within a range of $-10$ to $+40$ days. The successfully matched candidates are referred to as the ``TNS matched" sample.

Given that the redshift measurements for SN spectra by the DESI \texttt{redrock} pipeline may not always be accurate, we performed another cross-match similar to the one above, but this time on candidates identified prior to the visual inspection. The secondary cross-match imposed no restrictions on date or redshift, relying solely on position. Candidates matched in this manner that were not already in the ``TNS matched" sample are termed the ``TNS location matched" sample.

The remaining visually confirmed candidates are considered newly discovered SNe and are referred to as the ``New" sample. 

Because many transients in the TNS lack spectroscopic classifications (designated with the prefix "AT"), it is possible that the ``New" sample includes previously reported but unclassified targets. Therefore, we cross‑matched the Type Ia candidates in ``New" sample with the ATs discovered between 2020 November 30 and 2022 July 1. A search radius of 1 arcsec was used, and no restriction was placed on the time difference. 

\subsection{Type Ia supernovae}\label{subsec:result_Ia}
A total of 341 Type Ia candidates were automatically selected prior to visual inspection. After excluding candidates lacking distinct SN features or reliable spectral decompositions, 269 Type Ia SN candidates were identified in DR1, 22 of which were flagged for further analysis. Of the secure candidates, 60 have previously been classified on TNS, and 40 have been reported to the TNS but lack spectroscopic classifications. For one candidate with TARGETID 39628443943441091, two separate spectra taken at different epochs were identified; we treat these as independent entries in the following statistical analysis because \texttt{SNID} assigned them different classifications (Ia-norm and Ia-91T). Additionally, for a SN named 2021afii, its two best-matching templates are both Ic-norm, which contradicts its TNS classification (Ia-norm). Excluding the 22 ambiguous candidates and SN2021afii, the subtype distribution of the remaining 247 Type Ia SN spectra is presented in Table~\ref{type}, based on the type of the best‑matching template. Detailed information regarding these SNe is provided in Table~\ref{Tab:ALL}.

\begin{table}[htb]
\caption{Subtype distribution of 247 Type Ia SN spectra}
\begin{tabular}{lcc}
\hline
SN Ia subtype & Number & Fraction \\
\hline
Ia-norm       & 208     & 0.842     \\
Ia-91bg       & 5       & 0.020     \\
Ia-91T        & 25      & 0.101     \\
Ia-csm        & 0       & 0.000     \\
Ia-pec        & 7       & 0.028     \\
Ia-02cx       & 0       & 0.000     \\
Ia-03fg       & 1       & 0.004     \\
Ia-02es       & 1       & 0.004     \\
Ia-Ca-rich    & 0       & 0.000     \\
\hline
Total           & 247     & 1.000     \\
\hline
\end{tabular}
\tablecomments{Candidate 39628443943441091's two spectra are treated seperately}
\label{type}
\end{table}

\begin{figure}[htb!]
    \centering
    \gridline{
    \fig{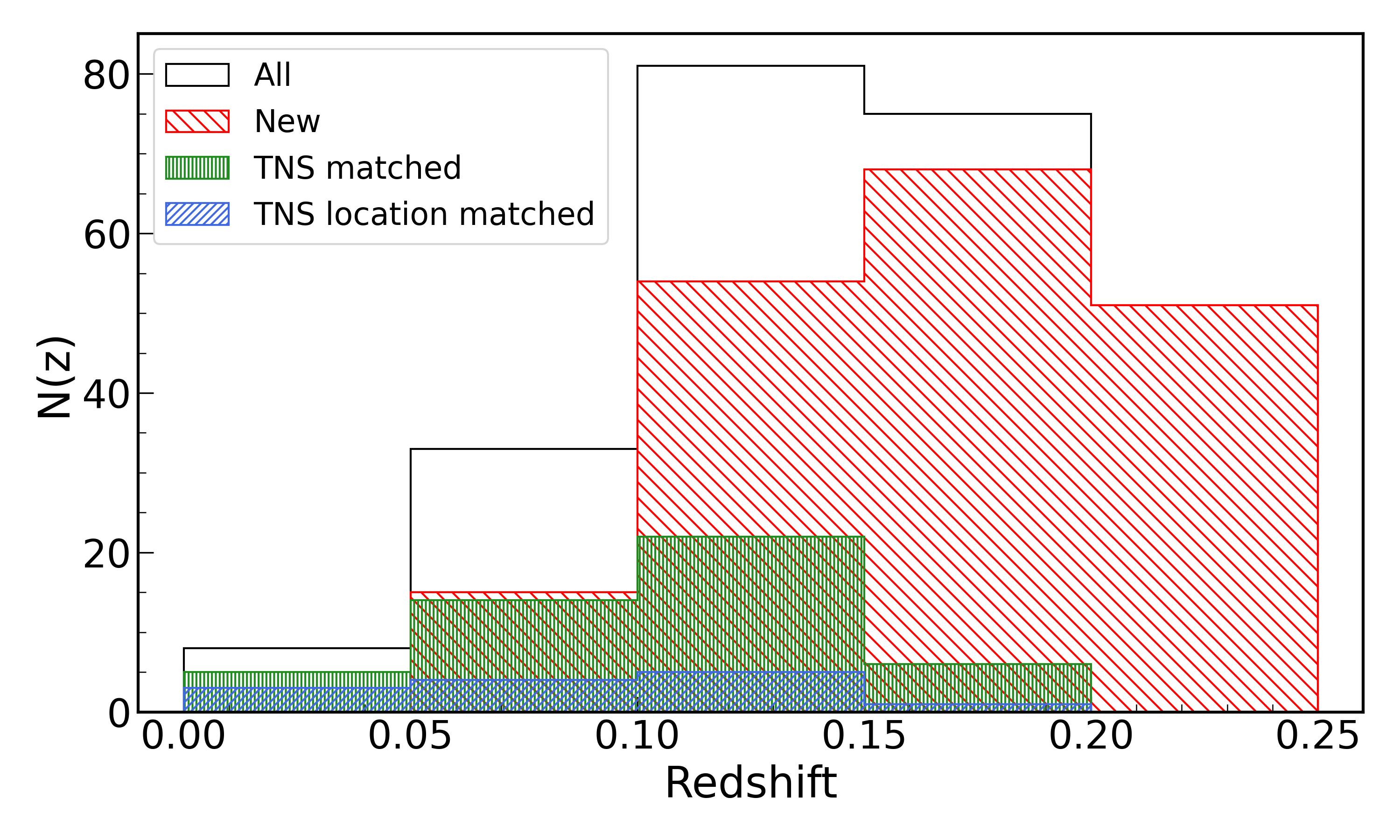}{0.46\textwidth}{(a)The redshift distribution of 247 Type Ia SNe}
    }
    \gridline{
    \fig{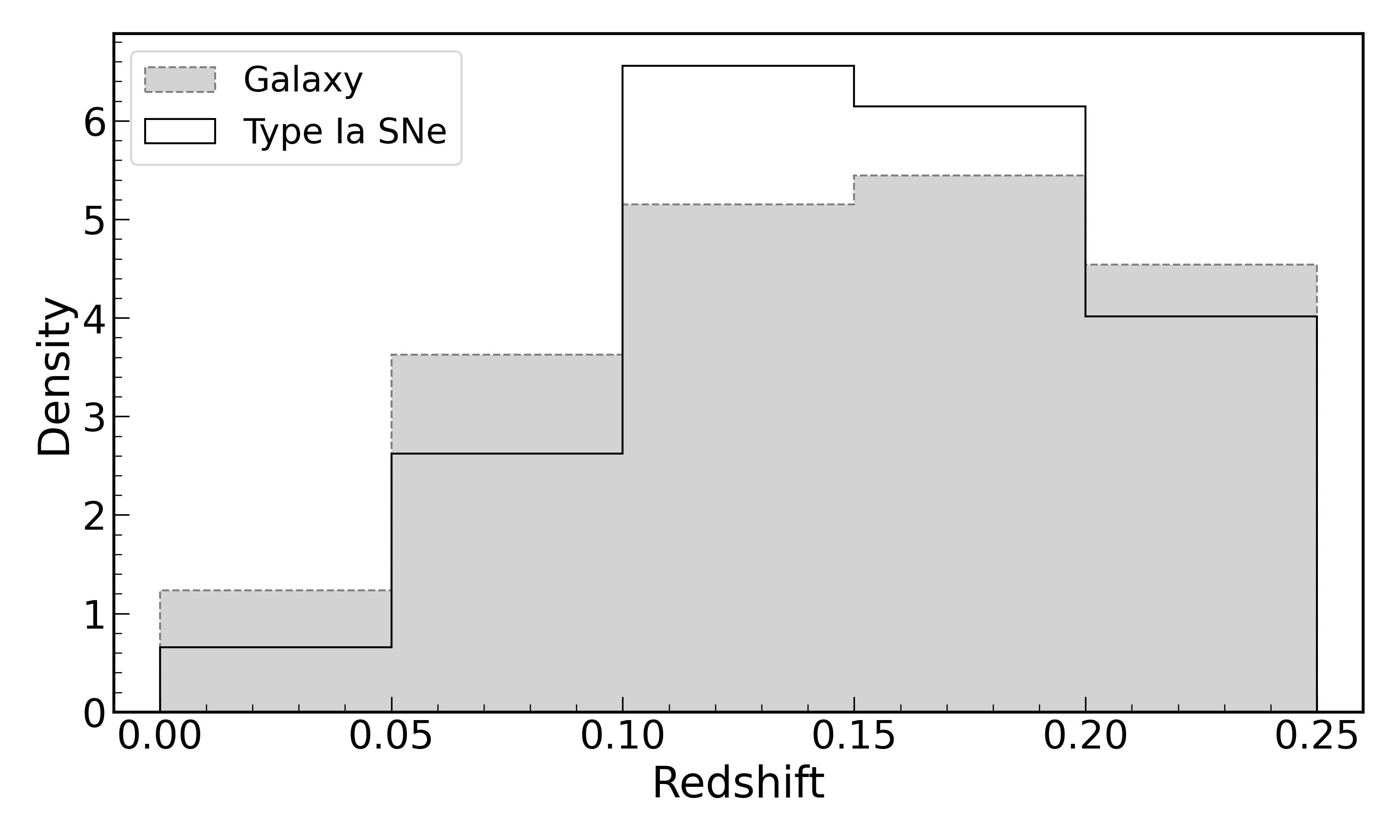}{0.46\textwidth}{(b)The redshift distribution comparison of all galaxies and SNe}
    }
    \caption{Redshift distribution of the samples. The redshift information comes from DESI DR1 catalog. ``TNS matched" sample (green) and ``TNS location matched" sample (blue) have a median $z=0.1075$, ``New" sample (red) has a median $z=0.1690$, all SNe (solid line) have a median $z=0.1513$, and all galaxies (dashed line) have a median $z=0.1499$. }
    \label{fig:zdis}
\end{figure}

The redshift distribution of the secure SNe and all sample galaxies is plotted in Figure~\ref{fig:zdis}. The figure reveals that Type Ia SNe previously cataloged on TNS are concentrated at redshifts of $z=0.05$--$0.15$, with a cutoff beyond $z=0.2$. In contrast, the ``New" SNe are concentrated in the range of $z=0.1$--$0.25$, with a median $z=0.169$. This difference demonstrates that the 4\,m Mayall telescope used by DESI reaches sufficient depths to detect SNe at intermediate redshifts, thereby enabling the discovery of transients missed by ordinary photometric surveys. It is likely that a considerable number of SNe could be found at $z > 0.25$; however, due to the wavelength limitations of the templates and rest-frame SN features, we restrict our sample to $z \le 0.25$.

\begin{figure*}[htbp]
    \centering
    \gridline{
        \fig{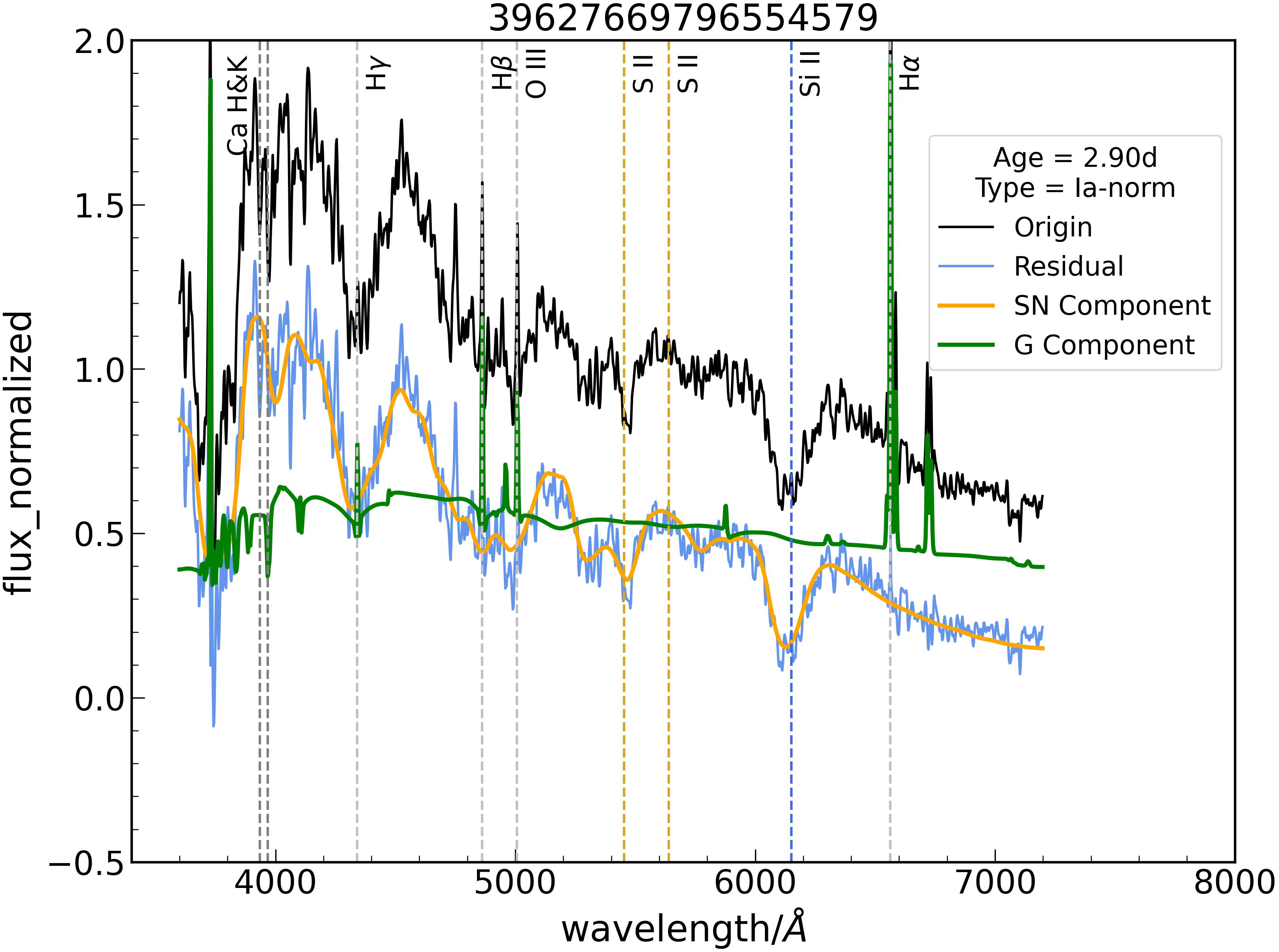}{0.44\textwidth}{}
        \fig{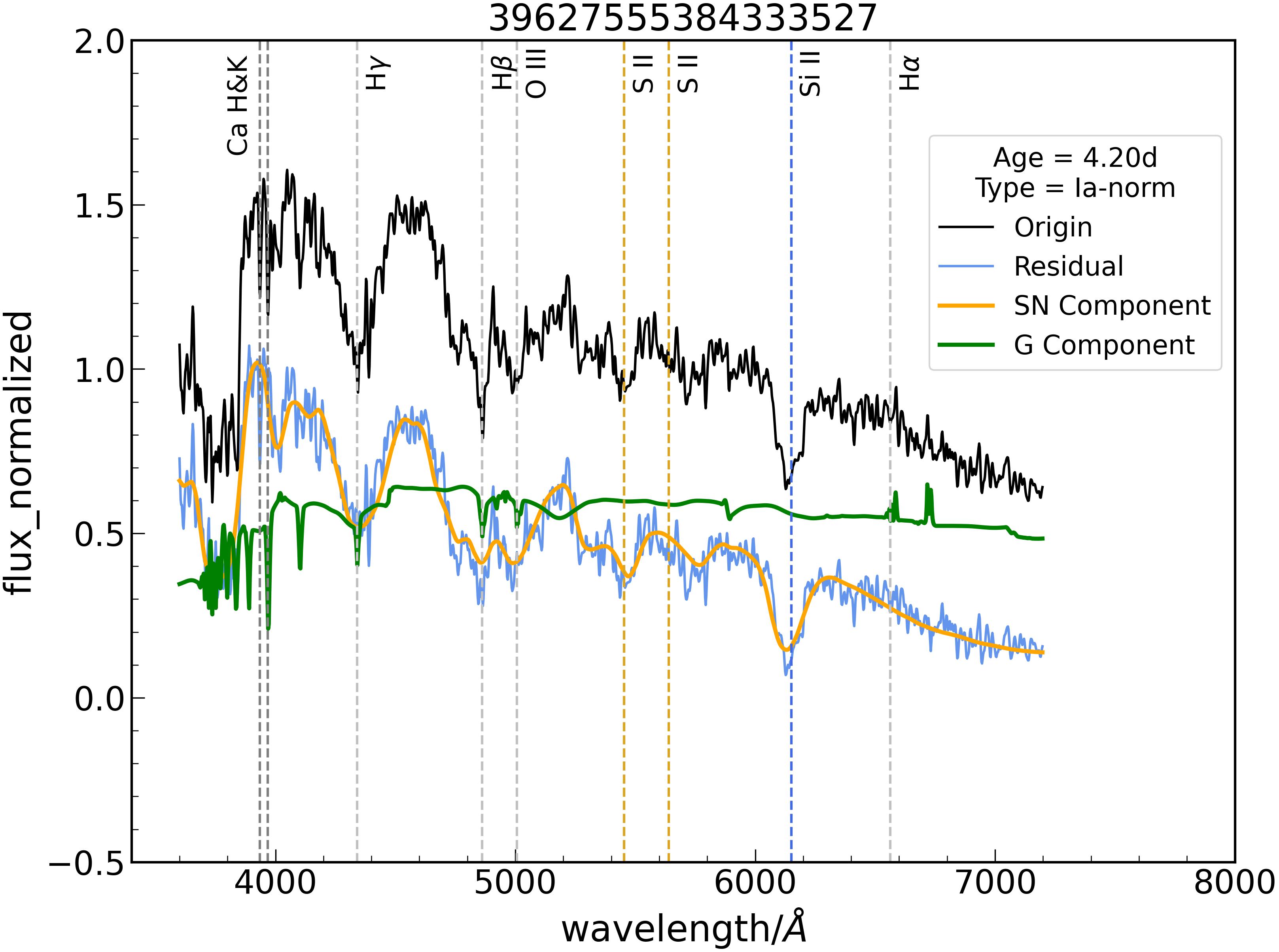}{0.44\textwidth}{}
    }
    \gridline{
        \fig{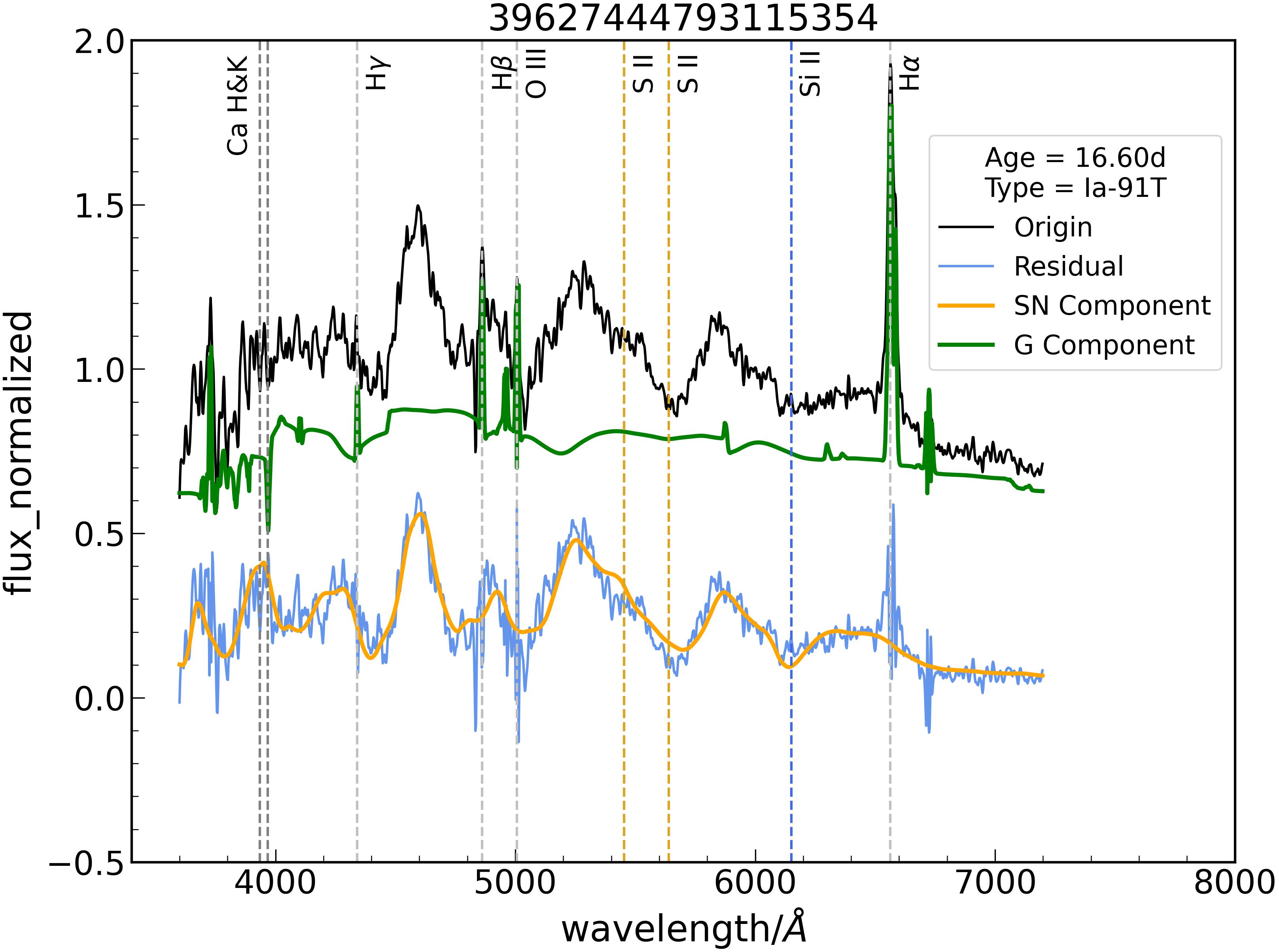}{0.44\textwidth}{}
        \fig{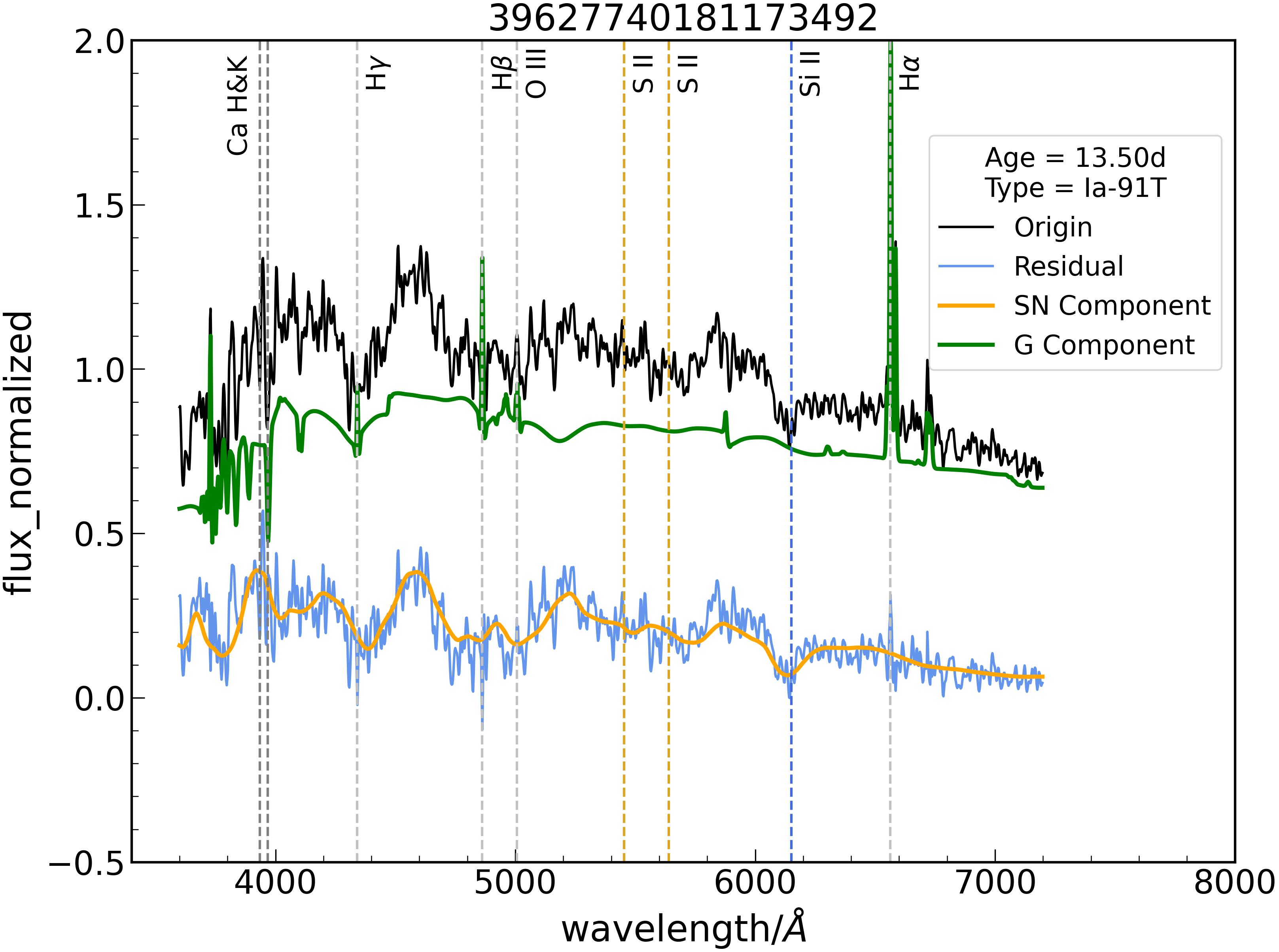}{0.44\textwidth}{}      
    }
    \gridline{
        \fig{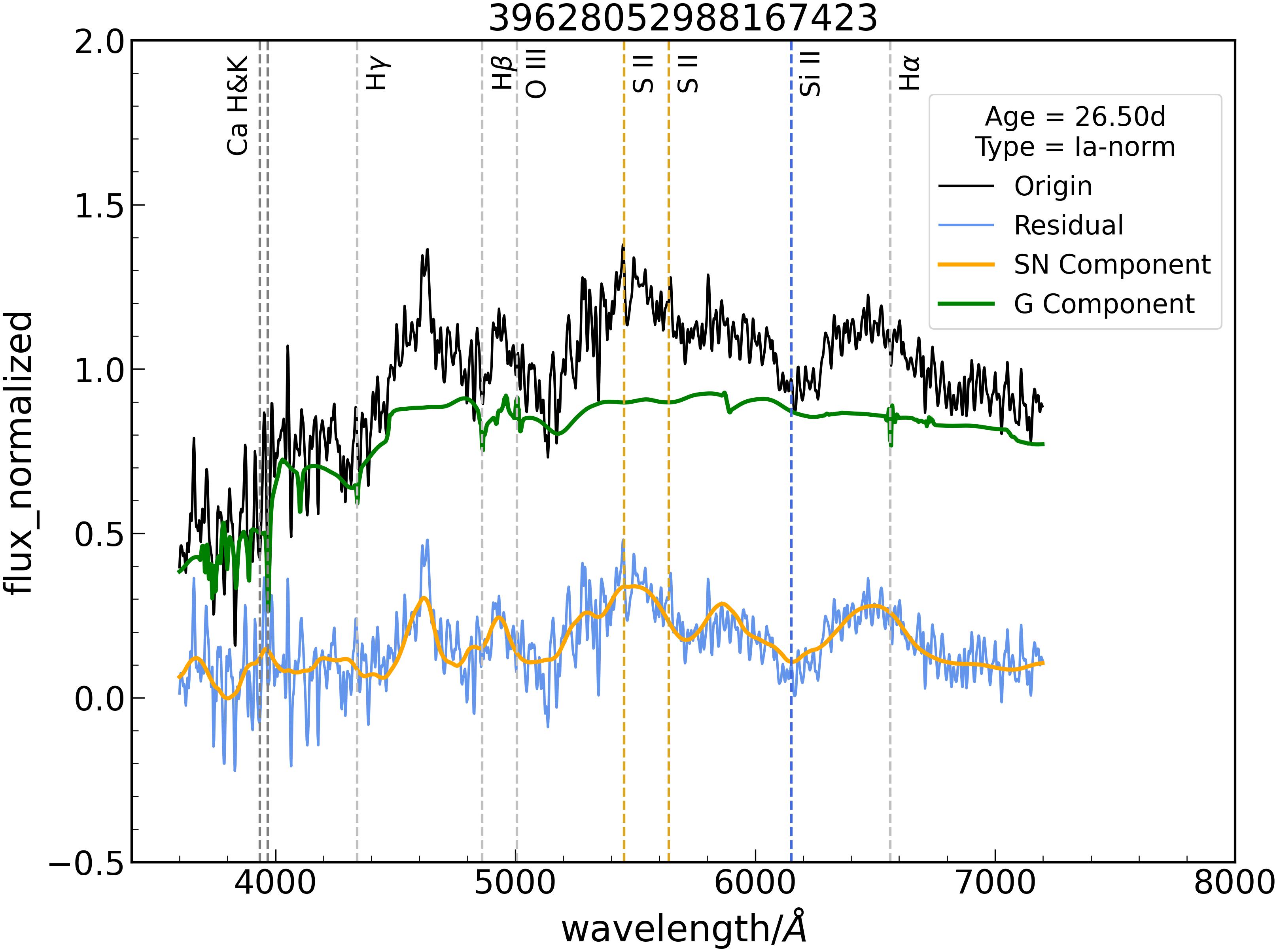}{0.44\textwidth}{}
        \fig{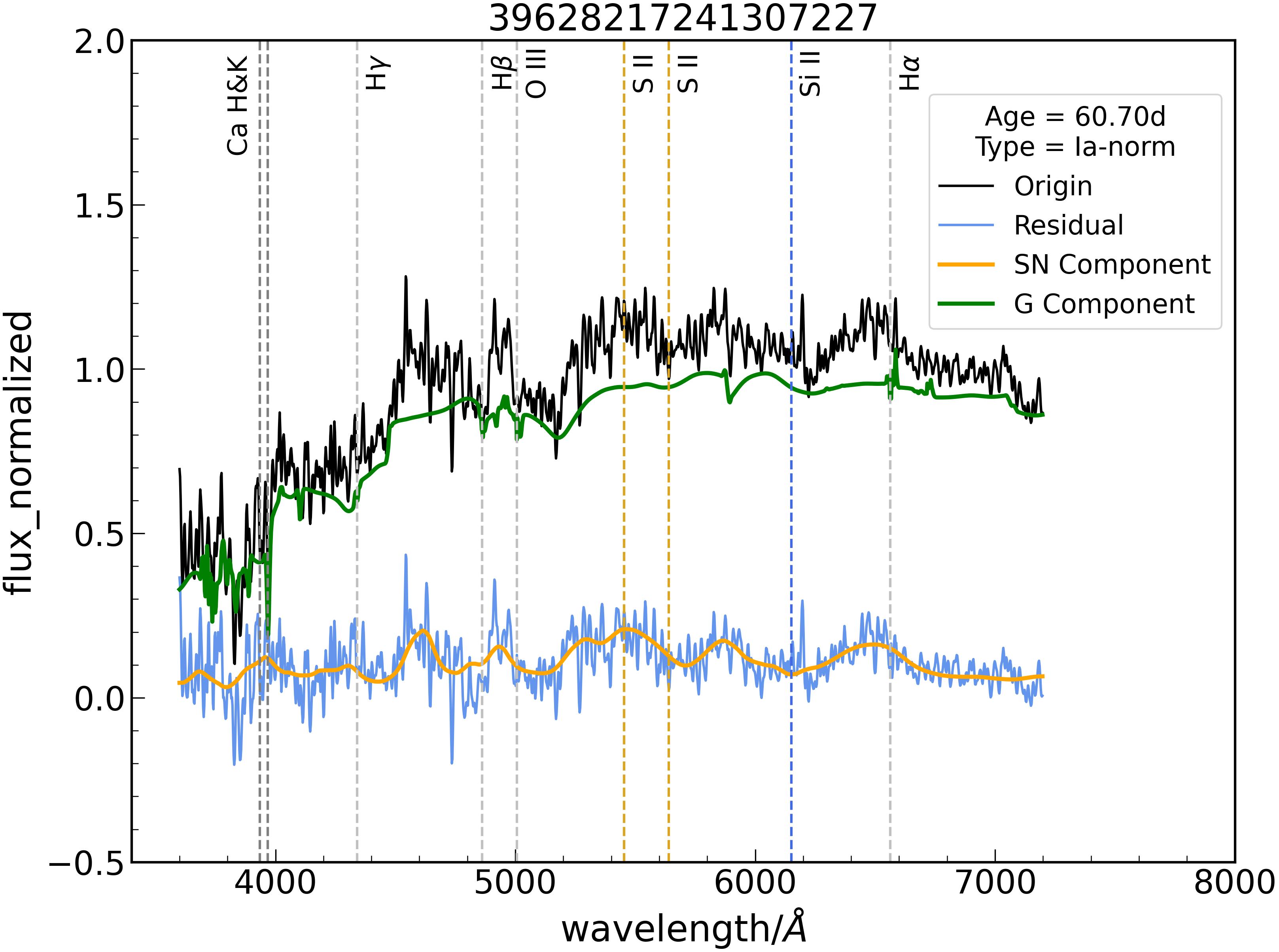}{0.44\textwidth}{}   
    }
    \caption{The spectra of 6 ``New" Type Ia SNe and their decomposition results. The spectra are plotted in rest frame and normalized by their median flux. The deep blue line is the processed spectrum, the green line is the galaxy component, the yellow line is the SN component, and the light blue line is the residual spectrum used for classification. Age and type are given by the best-matching \texttt{SNID} template. }
    \label{fig:IaSample}
\end{figure*}

\begin{figure*}[htb!]
    \centering
    \gridline{
        \fig{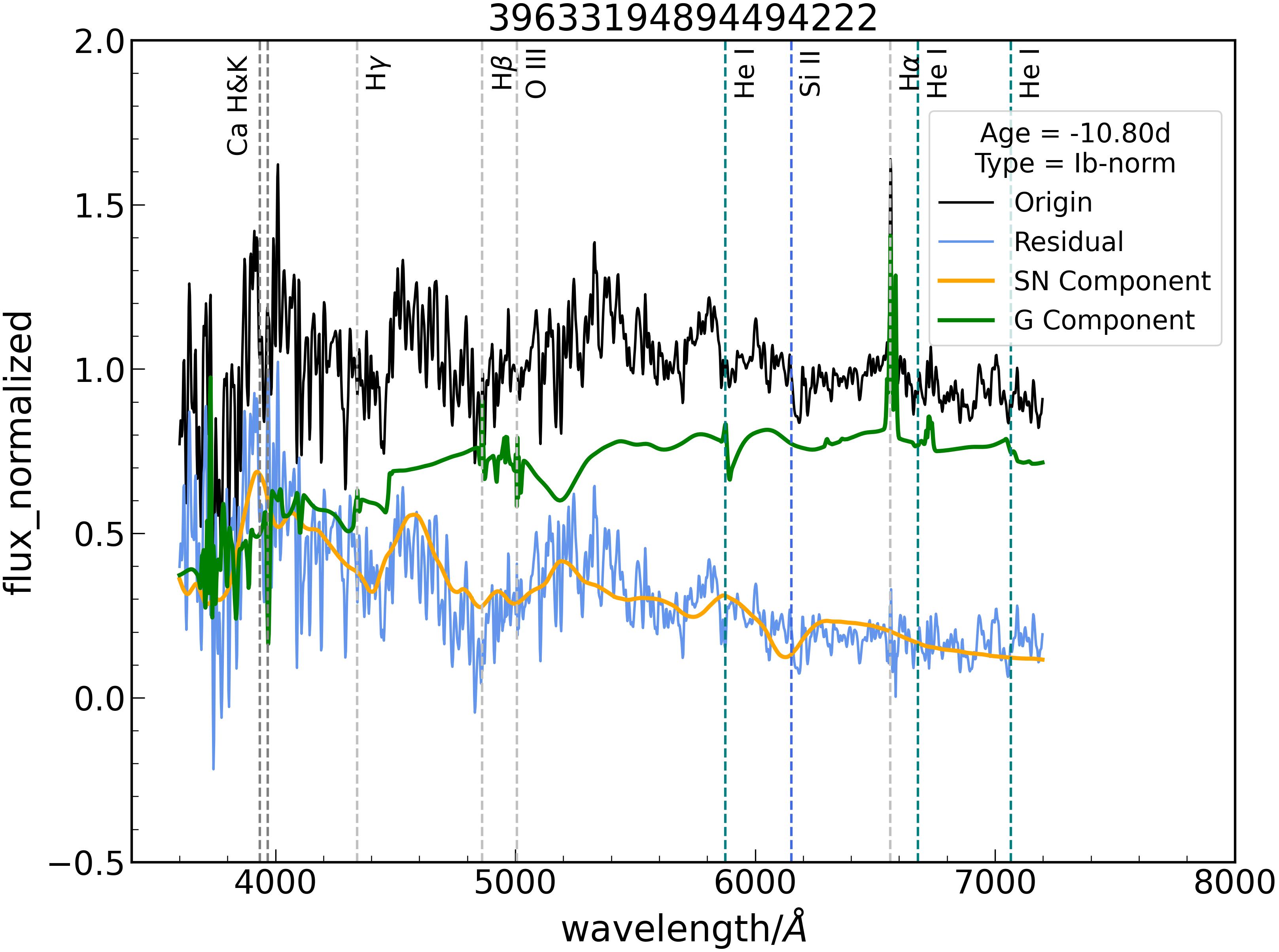}{0.44\textwidth}{}
        \fig{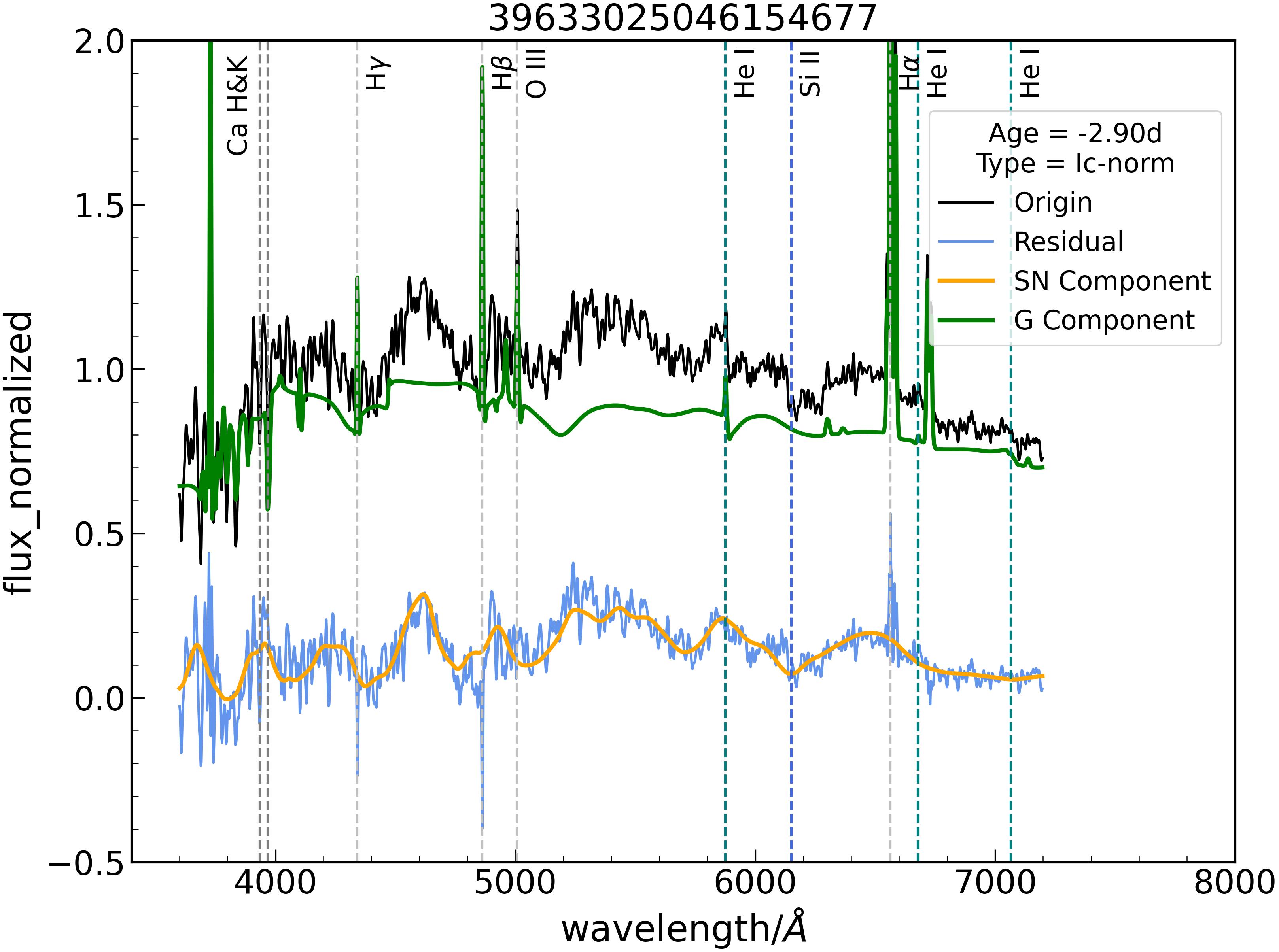}{0.44\textwidth}{}
    }
    \caption{The spectra of 1 Ib-norm SN (left) and 1 Ic-norm SN (right). The spectra are plotted in rest frame and normalized by the median flux. The deep blue line is the processed spectrum, the green line is the galaxy component, the yellow line is the SN component, and the light blue line is the residual spectrum used for classification. Age and type are given by the best-matching \texttt{SNID} template. }
    \label{fig:CCSample}
\end{figure*}

\begin{deluxetable*}{cccccccccc}
\label{Tab:ALL}
\tablecaption{Information of the 264 SNe found in DESI DR1 (partial)}
\tabletypesize{\footnotesize}
\tablehead{
\colhead{TNSname\tablenotemark{a}} & \colhead{TARGETID\tablenotemark{b}} & \colhead{z-TNS\tablenotemark{c}} & \colhead{z-DESI\tablenotemark{d}} & \colhead{RA[deg]\tablenotemark{e}} & \colhead{DEC[deg]\tablenotemark{f}} & \colhead{Date-TNS\tablenotemark{g}} & \colhead{Date-DESI\tablenotemark{h}} & \colhead{Type\tablenotemark{i}} & \colhead{Age[d]\tablenotemark{j}}
}
\startdata
SN2021ntw & 39628336003025098 & 0.099 & 0.092 & 259.6419 &  23.3413 & 20210529 & 20210618 & Ia-norm &  13.20 \\
SN2022gdq & 39633240440441234 & 0.090 & 0.093 & 199.3932 &  49.4134 & 20220403 & 20220414 & Ia-91T & 0.40 \\
--- & 39627669796554579 & --- & 0.175 & 345.0446 & -4.9340 & --- & 20211120 & Ia-norm & 2.90 \\
AT2021aadz & 39627444793115354 & --- & 0.153 & 61.5697 & -14.2985 & 20211001 & 20211026 & Ia-91T & 16.60 \\
--- & 39628217241307227 & --- & 0.228 & 263.2284 & 17.8970 & --- & 20210523 & Ia-norm & 60.70 \\
--- & 39627811912157419 & --- & 0.082 & 180.3834 & 0.9592 & --- & 20210502 & Ib-norm & -4.50 \\
--- & 39633025046154677 & --- & 0.186 & 194.6555 & 37.3185 & --- & 20220416 & Ic-norm & -2.90 \\
AT2021kcb & 39628443943441091 & --- & 0.148 & 194.4069 & 28.2547 & 20210417 & 20210505 & Ia-91T & 13.50 \\
AT2021kcb & 39628443943441091 & --- & 0.148 & 194.4069 & 28.2547 & 20210417 & 20210418 & Ia-norm & 6.88 \\
--- & 39627769679708563 & --- & 0.221 & 183.0145 & -0.7065 & --- & 20220430 & Ia-03fg & 2.00 \\
\enddata
\tablenotetext{a}{Corresponding object name in TNS}
\tablenotetext{b}{Unique DESI target ID}
\tablenotetext{c}{Redshift provided by TNS}
\tablenotetext{d}{Redshift provided by DESI}
\tablenotetext{e}{Barycentric Right Ascension in ICRS, provided by DESI}
\tablenotetext{f}{Barycentric Declination in ICRS, provided by DESI}
\tablenotetext{g}{Discovery Date provided by TNS, in YYYYMMDD}
\tablenotetext{h}{Final night of observation included in a series of spectra  coadds, in YYYYMMDD}
\tablenotetext{i}{SN type of the best-matching \texttt{SNID} template}
\tablenotetext{j}{SN age of the best-matching \texttt{SNID} template}
\tablecomments{Table~\ref{Tab:ALL} is published in its entirety in the machine-readable format. A portion is shown here for guidance regarding its form and content. }
\tablecomments{The 22 ambiguous candidates are not listed in the Table. }
\digitalasset
\end{deluxetable*}

\begin{figure*}[htb!]
    \centering
    \gridline{
    \fig{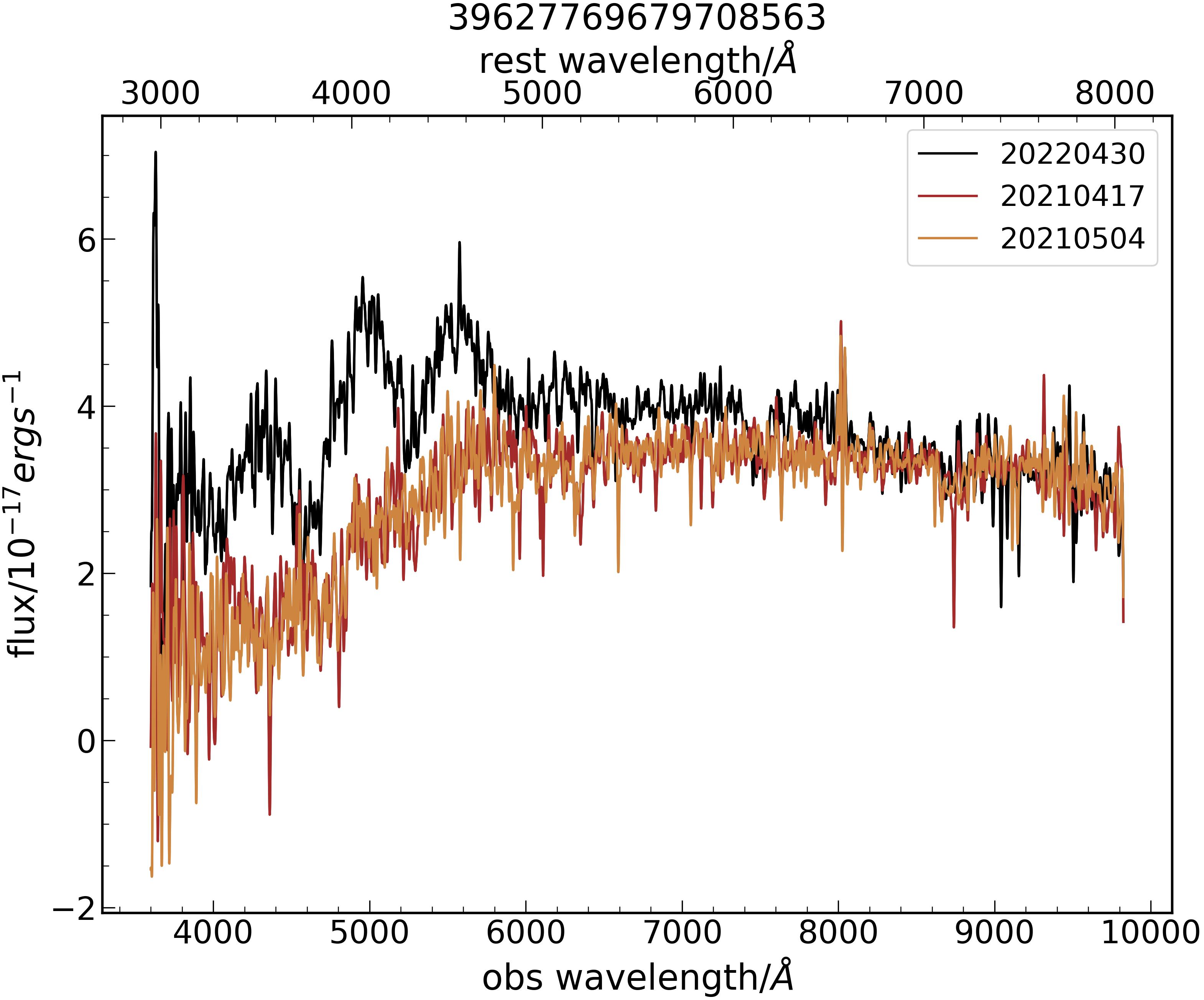}{0.46\textwidth}{(a)}
    \fig{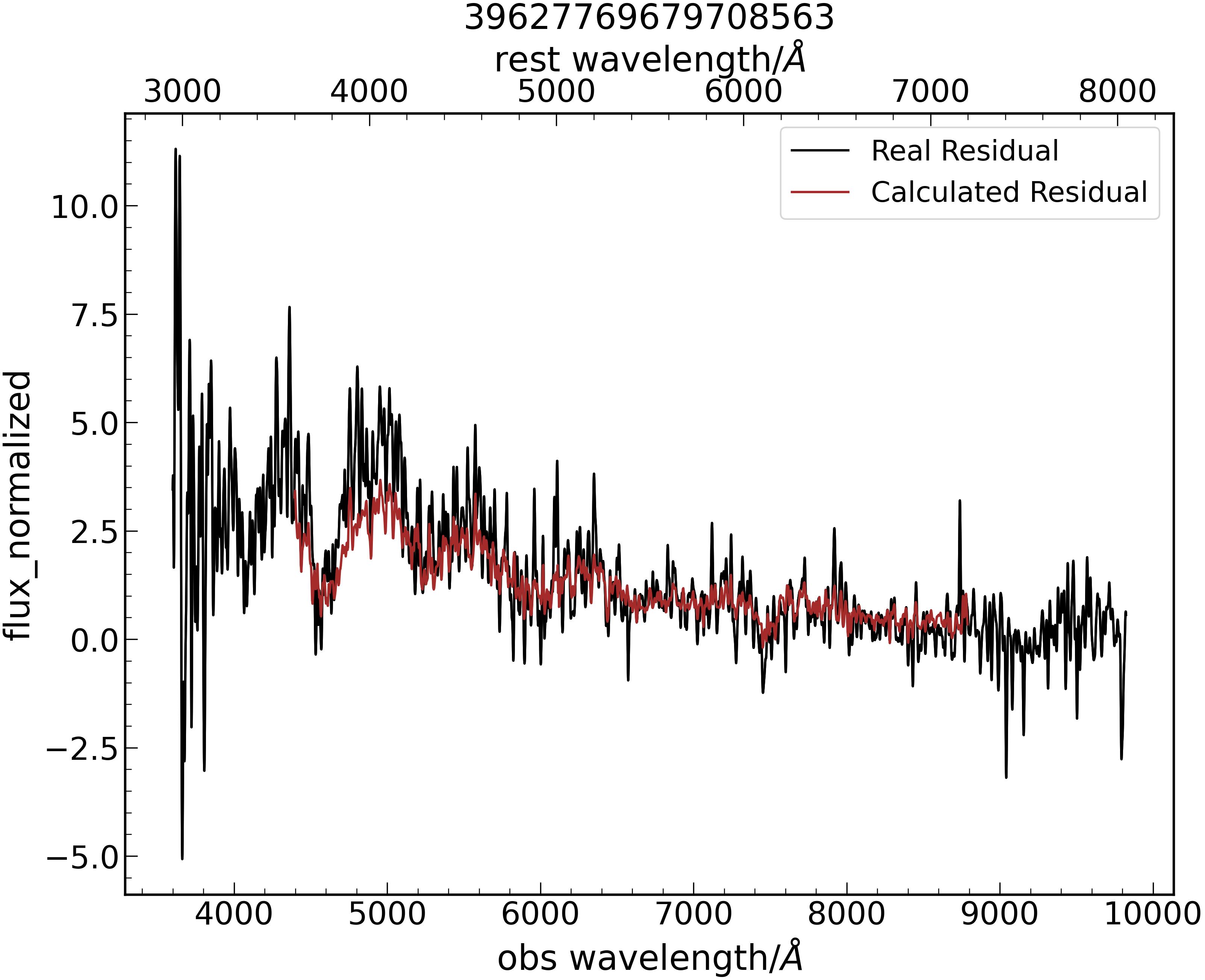}{0.46\textwidth}{(b)}
    }
    \gridline{
    \fig{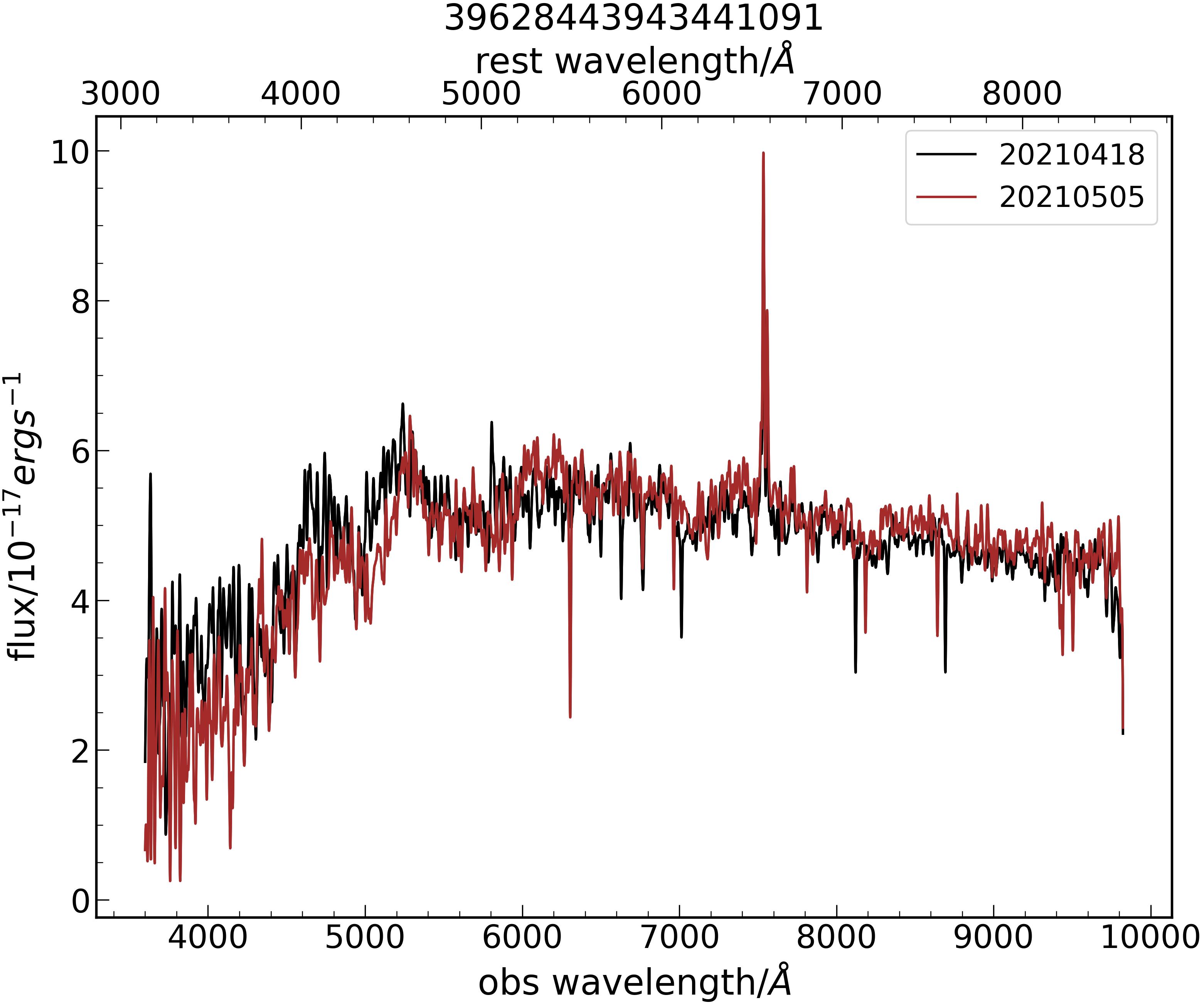}{0.46\textwidth}{(c)}
    \fig{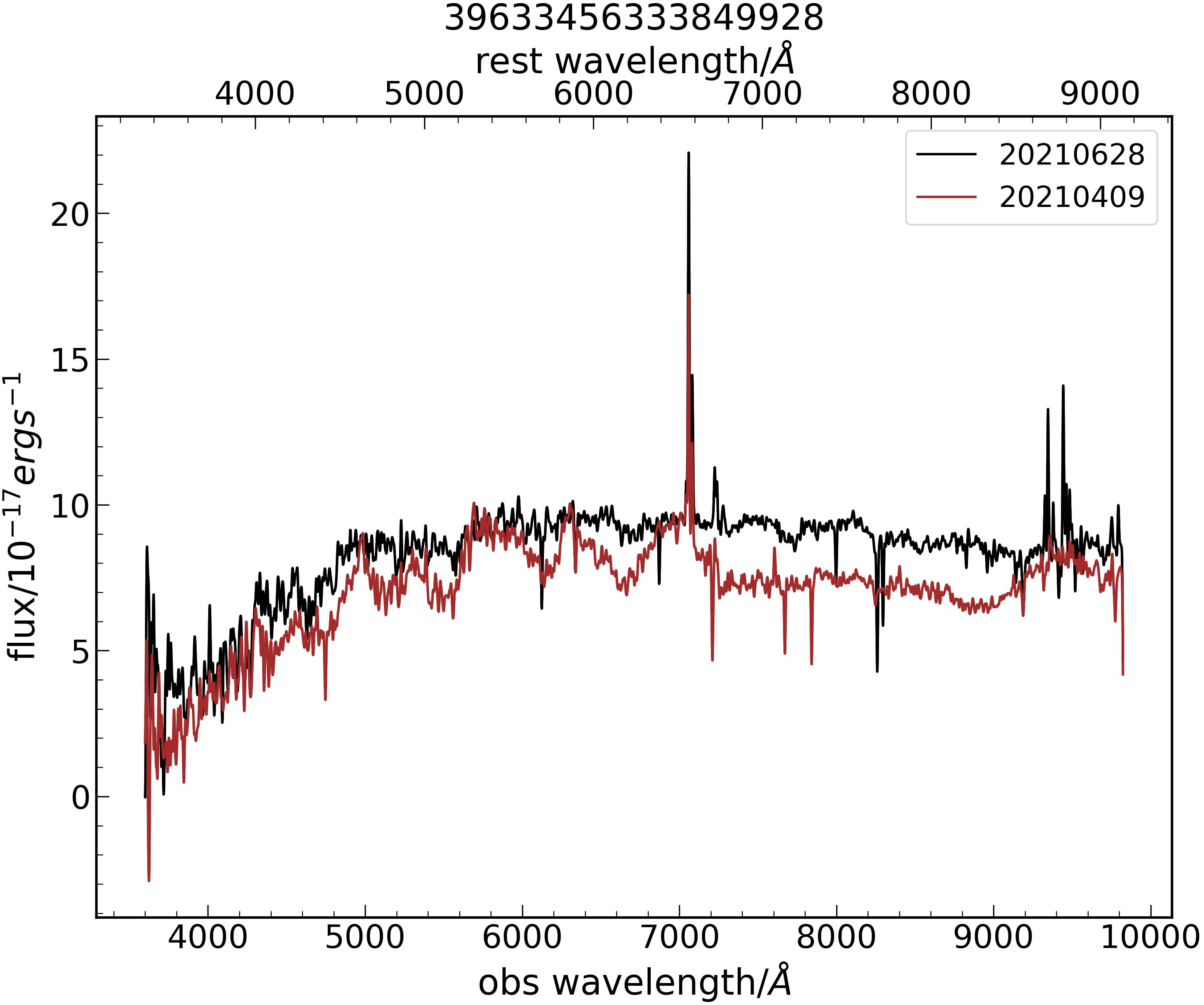}{0.46\textwidth}{(d)}
    }
    \caption{Candidates with multiple spectra in DR1. (a) Candidate having one spectrum exhibiting SN features and multiple host galaxy spectrum; (b) Comparison of the actual residual spectra (gray) with the residual spectra used for classification (red); (c) Candidate having spectra taken at different epochs of SN evolution; (d) The continuum flux of the candidate's host galaxy spectrum (black) exceeds that of the spectrum containing SN features (red). This is likely due to uncertainties in the flux calibration of the DESI pipeline, which is not optimized for spectra containing transient sources.}
    \label{fig:multi}
\end{figure*}

Figure~\ref{fig:IaSample} presents the spectra of six new SNe alongside their decomposition results. The method in Section \ref{sec:method} can successfully extract the SN component from spectra exhibiting SN feature. The galaxy component can mitigate the influence of narrow emission lines on the residual spectrum, improving the accuracy of correlation. 

\subsection{Supernovae of other types}

Since LOF solely identifies outliers and SNe share some common features, SNe of other types may also be selected despite defective decomposition. Before visual inspection, 274 non-Ia candidates were flagged, though many were misclassified due to poor decomposition or strong emission line contamination. Combining spectral features and information from the matched templates, we finally identified 17 SNe of other types in DR1, two of which were already recorded in the TNS. Specific information for these SNe can be found in Table \ref{Tab:ALL}.

\begin{figure*}[htb!]
    \centering
    \gridline{
    \fig{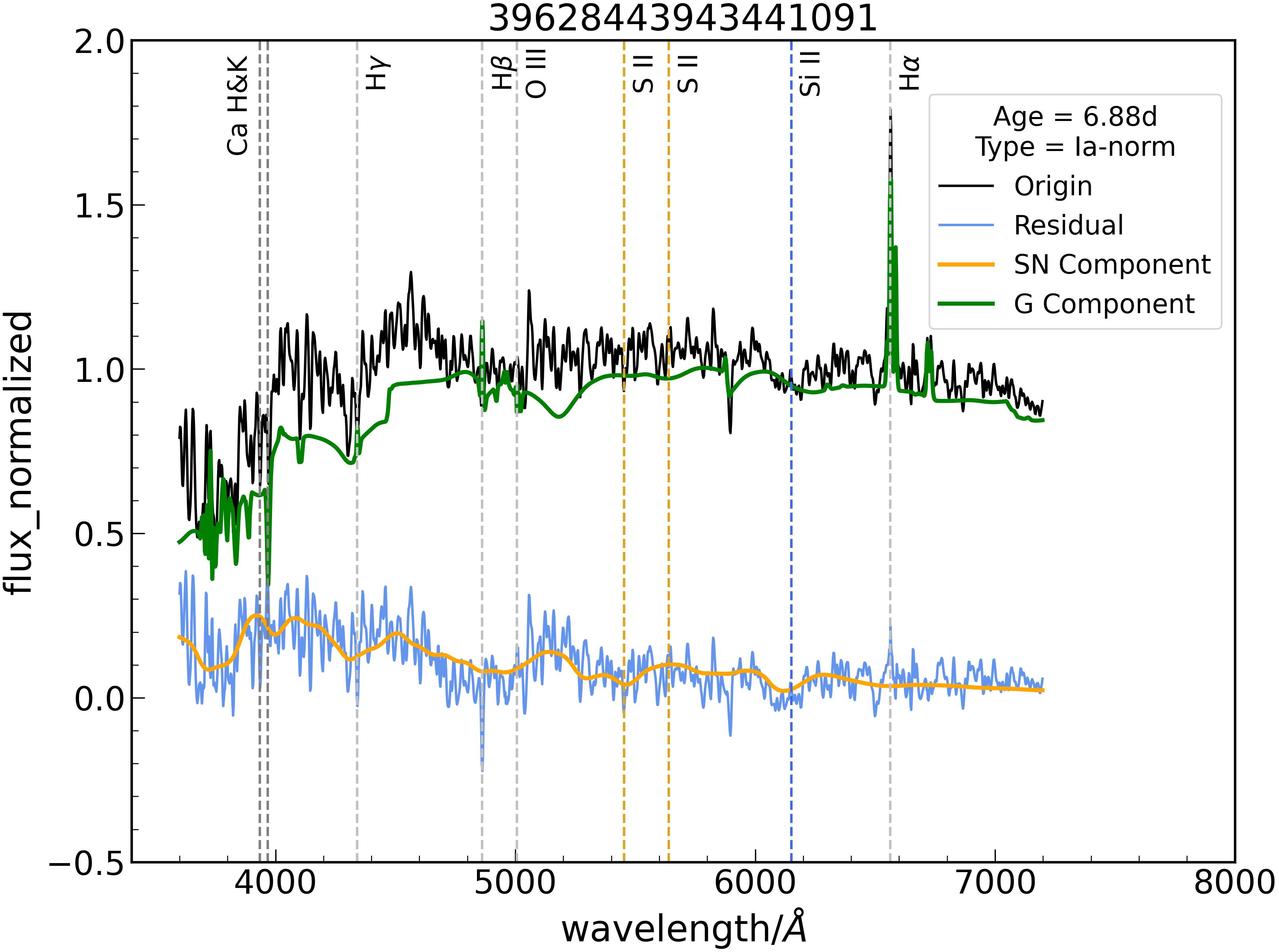}{0.46\textwidth}{(a) The early spectrum taken on 2021 April 18}
    \fig{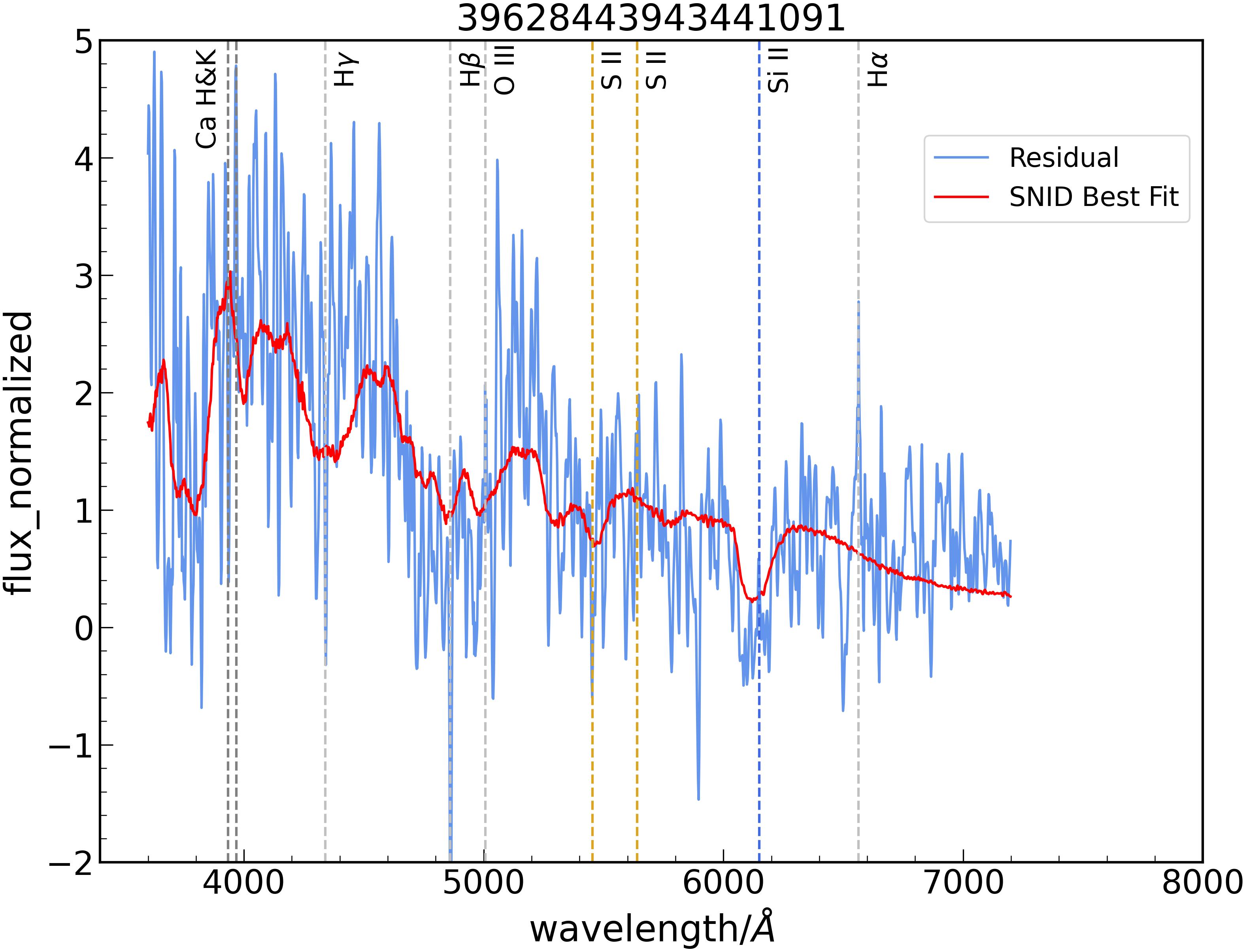}{0.46\textwidth}{(b) The residual of the early spectrum and its best-matching \texttt{SNID} template. }
    }
    \gridline{
    \fig{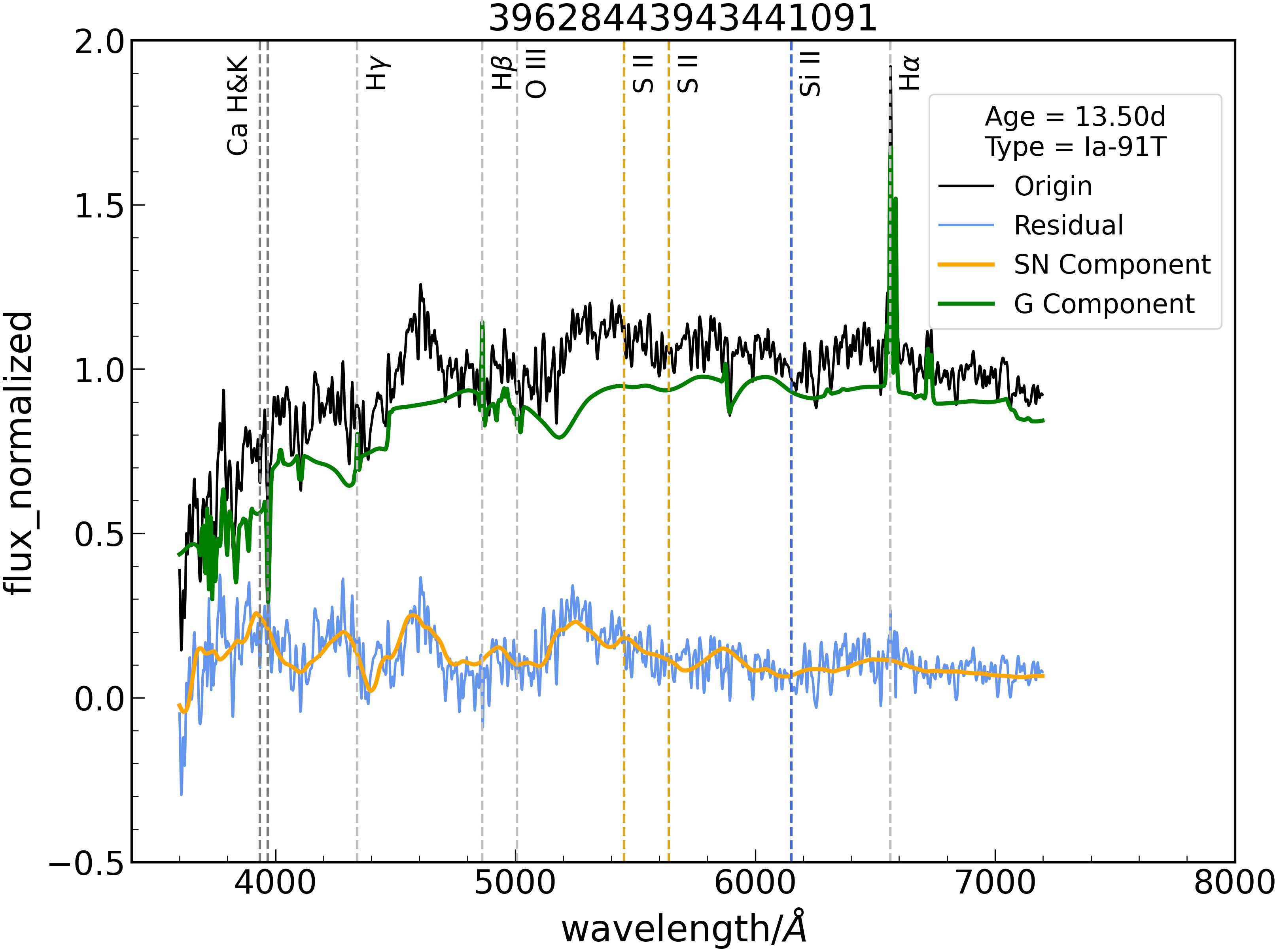}{0.46\textwidth}{(c) The late spectrum taken on 2021 May 5}
    \fig{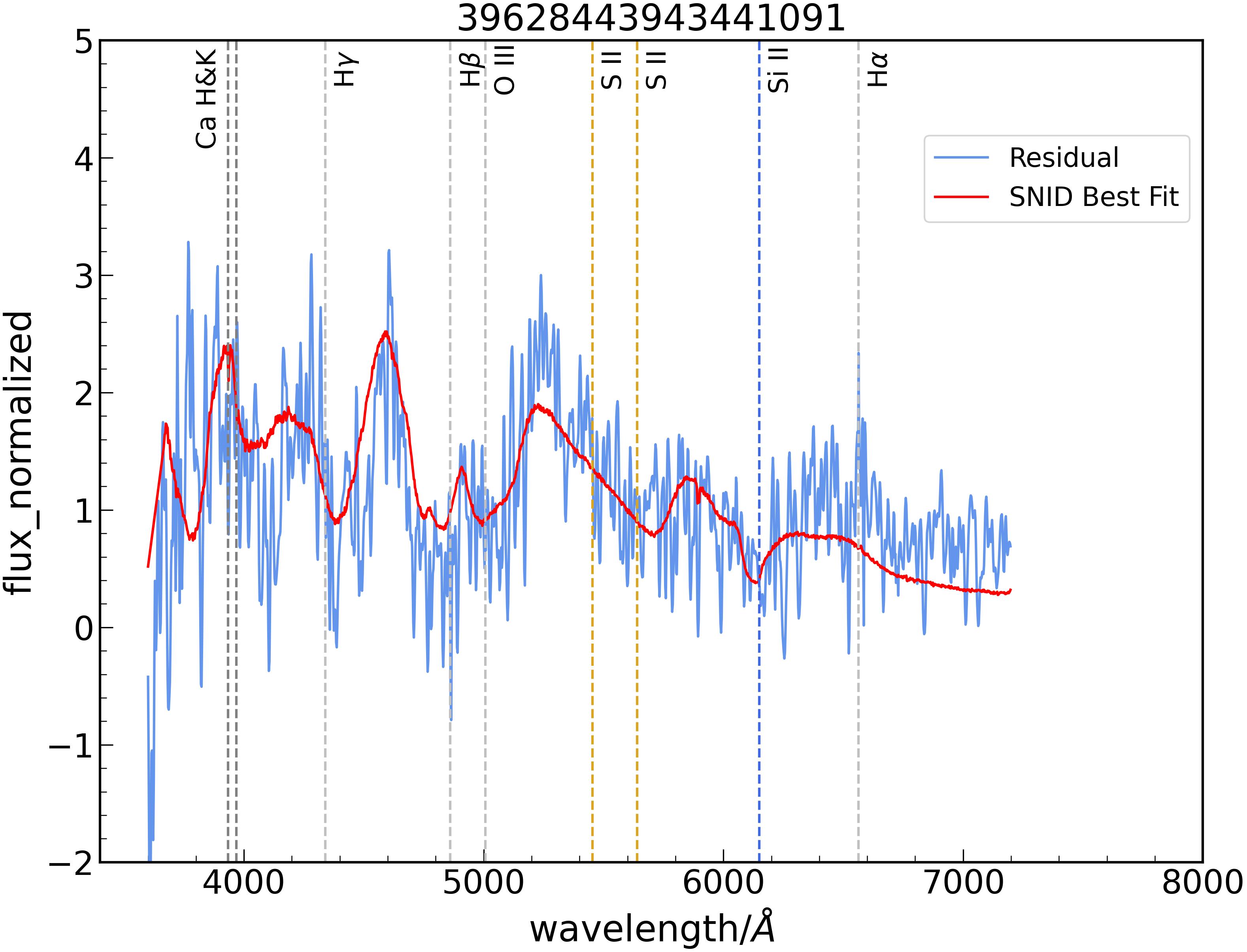}{0.46\textwidth}{(d) The residual of the late spectrum and its best-matching \texttt{SNID} template. }
    }
    \caption{The two spectra of 39628443943441091 and their decomposition result. Evolution do exist in the two spectra, but the SN features are too weak to do further classifications. A contrast of those two spectra is available in Figure \ref{fig:multi}. }
    \label{fig:AT2021kcb}
\end{figure*}

\subsection{Supernovae with multiple spectra}

A total of 15 candidates have multiple spectra in DR1, two of which are ambiguous candidates. Most candidates have one spectrum exhibiting SN features and one host galaxy spectrum, while some candidates possess spectra taken at different epochs of the SN evolution (see Figure~\ref{fig:multi}~(c)). 

For SN 2022jdw, two spectra with different TARGETIDs were cross-matched with it. The spectrum exhibiting SN features was successfully identified, while the other spectrum is presumed to be its host galaxy taken at a different pointing.

For the confirmed SNe, we subtracted the host galaxy component from the spectra exhibiting SN features and compared the results with the residual spectra used for classification in Section \ref{subsec:class}, as shown in Figure~\ref{fig:multi}~(b). This comparison demonstrates that the residual spectra adopted for classification are in excellent agreement with the actual residual spectra.

For some candidates, the continuum flux of the host galaxy spectrum exceeds that of the spectrum containing SN features, as seen in Figure~\ref{fig:multi}~(d). This may be due to issues with the flux calibration in DESI pipeline. This also indicates that the SN spectra in DESI may require more precise examination when used in flux-related analysis.

\section{Discussion}\label{discuss}
\subsection{Type Ia SN with different \texttt{SNID} classifications}

As noted in Section \ref{subsec:result_Ia}, the candidate 39628443943441091 possesses two spectra assigned different SN Ia subtypes. This candidate spatially matches the TNS transient AT 2021kcb, which was discovered on 2021 April 17. However, AT2021kcb lacks light curve data, making it difficult to determine its luminosity peak and and the exact phases of the two spectra. 

The earlier spectrum (2021 April 18) is classified by \texttt{SNID} as Ia-norm. Based on the discovery date and the observation date, this spectrum is likely pre-maximum. The lines of intermediate-mass elements (IMEs) such as Si~{\sc ii} and Ca~{\sc ii} can be seen in Figure \ref{fig:AT2021kcb} (b). If it is Ia-91T, this spectrum has already passed the early pre-maximum stage characterized by an almost featureless blue pseudo-continuum and the Fe~{\sc iii} absorption lines, which are the primary distinguishing features between Ia-norm and Ia-91T \citep{Mazzali_1995}. The later spectrum (2021 May 5) is classified by \texttt{SNID} as Ia-91T. This observation is likely post-maximum, as the peak of the blackbody continuum has shifted toward longer wavelengths. However, at post-maximum stage, Ia-91T can hardly be distinguished from Ia-norm \citep{Taubenberger_2017}.

Moreover, the shapes of S~{\sc ii} and Si~{\sc ii} are almost indistinguishable in Figure \ref{fig:AT2021kcb} (d), indicating that the SN features are significantly influenced by the host galaxy. The decomposition results also demonstrate that this SN did not significantly outshine its host galaxy. Subtracting the possibly overestimated galaxy component may change the intensities of certain absorption lines in the residual spectrum. Therefore, subtype classification based on specific absorption lines can be confusing. Limited by the spectral quality and phase uncertainties, we can only confidently classify AT 2021kcb as a Type Ia SN, treating the specific \texttt{SNID} subtypes as indicative rather than definitive.

\subsection{Unfound ``TNS matched" targets}

\begin{figure}[htb!]
    \centering
    \gridline{
        \fig{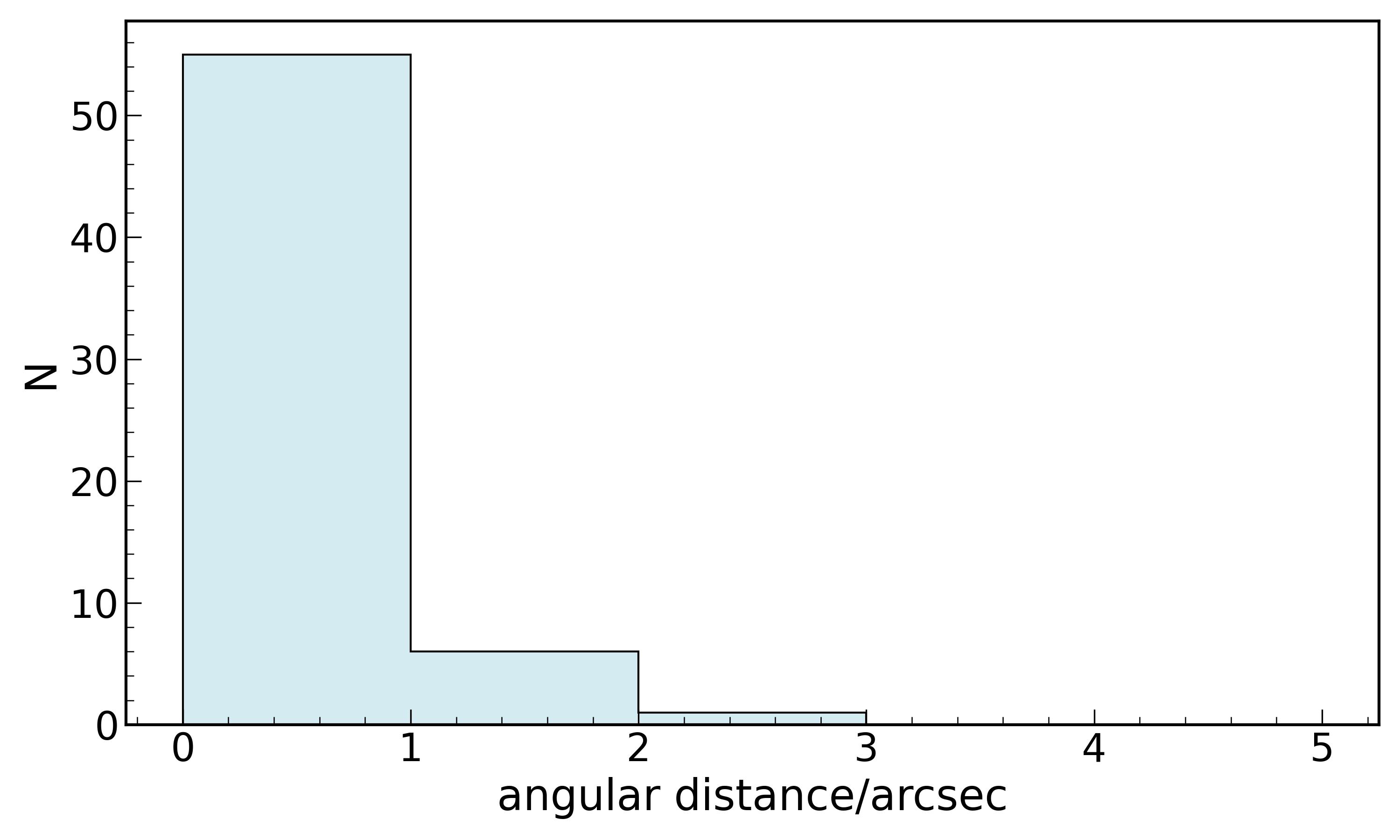}{0.46\textwidth}{(a)The angular separation distribution of 62 TNS SNe in candidates}
    }
    \gridline{
        \fig{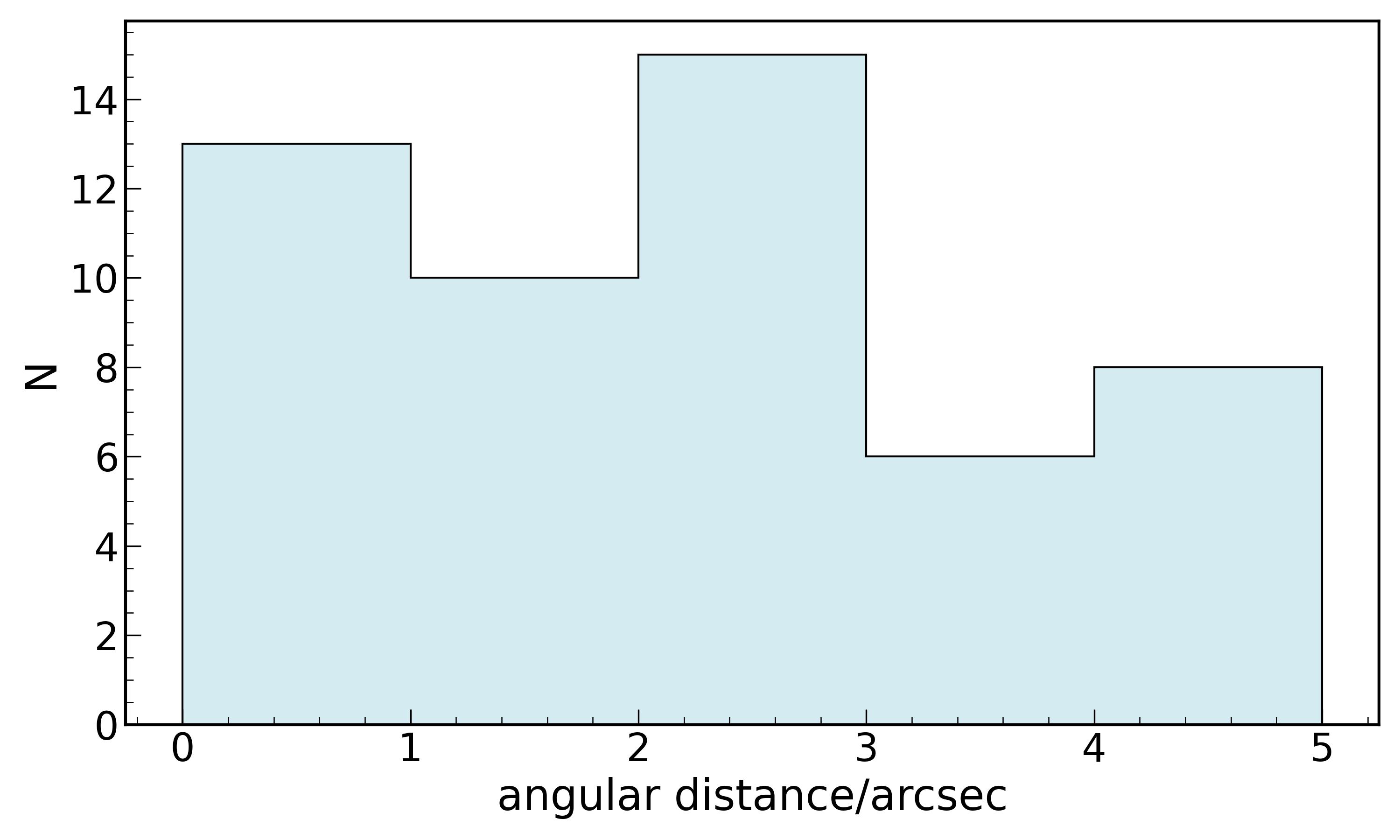}{0.46\textwidth}{(b)The angular separation distribution of 52 TNS SNe not in candidates}
    }
    \caption{The angular separation distribution of ``TNS matched" sample and ``TNS location matched" sample. Most of the found TNS SNe have angular distances less than 1 arcsec, while most of the not-found TNS SNe have larger angular distances. }
    \label{fig:angular}
\end{figure}

The ``TNS matched" sample was obtained by cross-matching all galaxy spectra, resulting in 49 SNe present in the SN candidates, and 52 not identified. By plotting the distribution of angular separation between TNS SNe and their matched DESI spectrum (as shown in Figure~\ref{fig:angular}), we found that the 49 ``TNS matched" objects in the candidates, along with 13 ``TNS location matched" objects, have very small angular separations (mostly less than 1 arcsec) from their corresponding DESI spectra. In contrast, the 52 SNe not in the candidates generally show larger angular separations from their matched DESI spectra. This suggests that the initial search radius used in the cross-match was too generous, leading to matches where the DESI fiber only captured the outer regions of the host galaxy, missing the actual SN flux entirely. 

We specifically examined the 13 targets not in the candidates that have angular separations smaller than 1 arcsec, to investigate why the algorithm failed to identify them. Among these, three are not Type Ia SNe; the poor decomposition of the Type Ia templates may have caused these SNe to blend in with normal galaxies. Three other SNe did pass the outlier detection, but were later discarded because of poor \texttt{SNID} matching results. For another three SNe, their spectra were taken nearly 10 days before the discovery date,meaning the observations likely predate the actual explosions. The remaining four SNe's corresponding spectra show no obvious SN features, despite being observed near their discovery dates. This is likely due to the intrinsic faintness of the SNe or slight pointing inaccuracies.

We also examined the 7 targets in the candidates that have angular separations greater than 1 arcsec. All of them have very obvious SN feature in their spectra, and the age estimated by \texttt{SNID} roughly agree with the difference between the observation date and the discovery date. One possible reason is that these SNe are bright enough to be detected even at distances greater than 1 arcsec. Another possible reason is the issues with pointing accuracy. 

\section{Summary and Outlook} \label{sec:sum}
We have conducted a supernova search using PCA and LOF on approximately 1.76 million spectra with $0<z<0.25$ from the DESI DR1. The candidates after LOF were then classified by \texttt{SNID} and confirmed via visual inspection. The search resulted in the discovery of 247 Type Ia SNe (and 22 ambiguous Type Ia candidates) and 17 SNe of other types. Among these, 187 Type Ia SNe and 15 other type SNe have no prior classification records in TNS. The SN sample we have constructed can provide a substantial number of high-resolution spectra for future astrophysical research and help to increase the number of SNe at the intermediate redshifts. The result also validates the feasibility of using DESI data as a powerful supplement source of SN spectra.

Future work on this topic will likely proceed in two directions. One direction involves utilizing the SN sample established in this work for further research to statistically explore the relationship between the SNe and their host galaxies. The other direction entails seeking suitable templates for higher redshifts to apply the searching method on higher-redshift spectra, thereby extending the SN search to galaxy spectra with redshifts above 0.25 and supplementing the high-redshift SN sample. Additionally, one could consider incorporating the latest machine learning algorithms to optimize the searching method, improving the efficiency, completeness and degree of automation of the procedure.

%% Please use the acknowledgment and contribution environments. This will 
%% be anonomyized when the "anonymous" style option is used. 
\begin{acknowledgments}

The authors would like to thank Prof. Hongxin Zhang, Prof. Teng Liu, Prof. Enci Wang, Prof. Ruizhi Yang and Assoc. Prof. Bo Wang for their helpful suggestions. The authors acknowledge the use of the DeepSeek-V3 and Gemini 3.1 language model for language editing 
assistance. This work is  supported by the National Science Foundation of China (NSFC, Grant No. 12233008, 12473008), the National Key R\&D Program of China (2023YFA1608100, 2025YFF0511003), the Strategic Priority Research Program of the Chinese Academy of Sciences (Grant No. XDB0550200), the Cyrus Chun Ying Tang Foundations, the 111 Project for "Observational and Theoretical Research on Dark Matter and Dark Energy" (B23042), and the Undergraduate Research Training Program of the University of Science and Technology of China (Project No. 20231102).

This research used data obtained with the Dark Energy Spectroscopic Instrument (DESI). DESI construction and operations is managed by the Lawrence Berkeley National Laboratory. This material is based upon work supported by the U.S. Department of Energy, Office of Science, Office of High-Energy Physics, under Contract No. DE–AC02–05CH11231, and by the National Energy Research Scientific Computing Center, a DOE Office of Science User Facility under the same contract. Additional support for DESI was provided by the U.S. National Science Foundation (NSF), Division of Astronomical Sciences under Contract No. AST-0950945 to the NSF’s National Optical-Infrared Astronomy Research Laboratory; the Science and Technology Facilities Council of the United Kingdom; the Gordon and Betty Moore Foundation; the Heising-Simons Foundation; the French Alternative Energies and Atomic Energy Commission (CEA); the National Council of Humanities, Science and Technology of Mexico (CONAHCYT); the Ministry of Science and Innovation of Spain (MICINN), and by the DESI Member Institutions: www.desi.lbl.gov/collaborating-institutions. The DESI collaboration is honored to be permitted to conduct scientific research on I’oligam Du’ag (Kitt Peak), a mountain with particular significance to the Tohono O’odham Nation. Any opinions, findings, and conclusions or recommendations expressed in this material are those of the author(s) and do not necessarily reflect the views of the U.S. National Science Foundation, the U.S. Department of Energy, or any of the listed funding agencies.
\end{acknowledgments}

\facility{Mayall}

\software{Astropy \citep{2013A&A...558A..33A,2018AJ....156..123A,2022ApJ...935..167A}, SNID \citep{Blondin_2007}, SNCosmo \citep{barbary_2025_15019859}, scikit-learn \citep{scikit-learn}, Matplotlib \citep{matplotlib}, NumPy \citep{numpy}, SciPy \citep{scipy}, SymPy \citep{sympy}, Pandas \citep{pandas}}


%% For this sample we use BibTeX plus aasjournalv7.bst to generate the

\bibliography{reference}{}
\bibliographystyle{aasjournalv7}



\end{document}